\documentclass[preprint]{aastex631}

\usepackage{amsmath}
\usepackage{float}
\received{June 8, 2023}
\submitjournal{APJ}

\shorttitle{AASTeX v6.3.1 Sample article}
\shortauthors{Rosborough et al.}
\graphicspath{{./}{figures/}}
\begin{document}

\title{Modeling the Reverberation Response of the Broad Line Region in Active Galactic Nuclei}

\author[0000-0001-5055-507X]{Sara A. Rosborough}
\affiliation{Laboratory for Multiwavelength Astrophysics, School of Physics and Astronomy, Rochester Institute of Technology,
84 Lomb Memorial Dr,
Rochester, NY 14623, USA}

\author[0000-0003-0672-2154]{Andrew Robinson}
\affiliation{Laboratory for Multiwavelength Astrophysics, School of Physics and Astronomy, Rochester Institute of Technology,
84 Lomb Memorial Dr,
Rochester, NY 14623, USA}

\author[0000-0001-7259-7043]{Triana Almeyda}
\affiliation{Department of Astronomy, University of Florida,
211 Bryant Space Science Center Stadium Road,
Gainesville, FL 32611, USA}

\author{Madison Noll}
\affiliation{Churchville-Chili Senior High School,
 5786 Buffalo Rd
 Churchville, NY 14428, USA}
\affiliation{Research Intern, Laboratory for Multiwavelength Astrophysics, School of Physics and Astronomy, Rochester Institute of Technology,
84 Lomb Memorial Dr 
Rochester, NY 14623, USA}



\begin{abstract} 
   
The variable continuum emission of an active galactic nucleus (AGN) produces corresponding responses in the broad emission lines, which are modulated by light travel delays, and contain information on the physical properties, structure, and kinematics of the emitting gas region.  The reverberation mapping technique, a time series analysis of the driving light curve and response, can recover some of this information, including the size and velocity field of the broad line region (BLR).  Here we introduce a new forward-modeling tool, the Broad Emission Line MApping Code (BELMAC), which simulates the velocity-resolved reverberation response of the BLR to any given input light curve by setting up a 3D ensemble of gas clouds for various specified geometries, velocity fields, and cloud properties.  In this work, we present numerical approximations to the transfer function by simulating the velocity-resolved responses to a single continuum pulse for sets of models representing a spherical BLR with a radiatively driven outflow and a disk-like BLR with Keplerian rotation. We explore how the structure, velocity field, and other BLR properties affect the transfer function.  We calculate the response-weighted time delay (reverberation ``lag''), which is considered to be a proxy for the luminosity-weighted radius of the BLR.  We investigate the effects of anisotropic cloud emission and matter-bounded (completely ionized) clouds and find the response-weighted delay is only equivalent to the luminosity-weighted radius when clouds emit isotropically and are radiation-bounded (partially ionized).  Otherwise, the luminosity-weighted radius can be overestimated by up to a factor of 2.  % 246/250 words  

\end{abstract}



\section{Introduction} \label{sec:intro}

Rapidly growing supermassive black holes (SMBHs) are fed by a surrounding accretion disk, which together compose the central engines of active galactic nuclei (AGN).  Most, if not all, large galaxies host a SMBH and its mass, $M_\bullet$, is closely tied to the galaxy's evolution.  Decades of $M_\bullet$ estimates show a correlation to the host galaxy's stellar velocity dispersion ($\sigma_\star$), the $M_\bullet-\sigma_\star$ relation \citep{McConnell2013REVISITINGPROPERTIES}, implying co-evolution between the SMBH and the bulge. The power of an AGN is regulated by the SMBH accretion rate, which is limited by its Eddington luminosity, where $L_{Edd} \propto M_\bullet$ and $M_\bullet$ can be $10^6-10^{10}$\, solar masses ($\,M_\odot$). Outflows and jets from AGN are thought to have a profound influence on galaxy evolution by suppressing star formation.  Therefore, establishing how the SMBH mass function evolves with redshift and luminosity is critical to understanding galaxy formation and evolution \citep{DiMatteo2005EnergyGalaxies,Hopkins2008AActivity, Fabian2012ObservationalFeedback,Alexander2012WhatHoles}.

\subsection{The Broad Line Region}

The AGN's central engine is surrounded by a geometrically and optically thick structure comprised of molecular gas and dust clouds, generally known as the dusty ``torus'' \citep[e.g.,][]{Urry1995UnifiedNuclei}.  Depending on grain composition and size, dust sublimates at temperatures between $\sim 1500-1800$\,K and cannot not survive the heating by the accretion disk's UV/optical continuum  within the corresponding dust sublimation radius, $R_d \sim 1$\,pc (assuming an AGN luminosity $\sim10^{45}$\,erg\,s$^{-1}$). Inside $R_d$ is the broad line region (BLR), a largely dust-free zone where dense gas clouds are photoionized by the continuum radiation \citep[e.g.,][]{Baskin2014RadiationNuclei,Baskin2018DustNuclei,Amorim2021A3783}. Residing well within the SMBH's gravitational sphere of influence, the gas clouds emit spectral lines that are Doppler-broadened by a several-tens of thousand km\,s$^{-1}$ \citep[e.g.,][]{Netzer2008IonizedNuclei}.  The BLR is therefore a unique diagnostic of the SMBH's mass and the physical processes that operate in its environment.  

The mass of the black hole, $M_{\bullet}$, can be estimated from the velocity dispersion ($\Delta v$) and radius ($R_{BLR}$) of the BLR,

\begin{equation}
\label{eq:BHmass}
   M_{\bullet} = f\frac{\Delta v^2 R_{BLR}}{G}
\end{equation}

where $G$ is the gravitational constant \citep[e.g.,][]{Bentz2015TheDatabase}.  However, since AGN are too distant and the BLR is far too small in angular size to obtain spatially resolved spectra, the BLR's structure and dynamics are not understood in  detail.  Therefore,    
%, $R_{BLR}$ is obtained from the reverberation time lag, and $\Delta v$ measured from the BEL profile's width.  
the virial factor, $f$, accounts for the unknown geometry and kinematics, and $\Delta v^2 R_{BLR}/G$ is the virial mass, $M_{\bullet,\,vir}$ \citep{Homayouni2020TheMonitoring,Pancoast2014ModellingResults,Netzer2010THENUCLEI}.  Comparisons of $M_{\bullet,\,vir}$ from the AGN that have been reverberation mapped to $M_{\bullet}$ determined with the $M_{\bullet} - \sigma_{\ast}$ scaling relationship indicate $f\approx4.5$ \citep{Woo2015THEGALAXIES,Batiste2017RecalibrationAGN}. 
 $\Delta v$ can be estimated from broad emission line (BEL) profile widths, but $R_{BLR}$ cannot be directly determined through imaging for most AGN (with a few exceptions discussed below).  Instead, $R_{BLR}$ can be determined by the reverberation mapping technique, which utilizes the response of the broad lines to the AGN's continuum variability and is not fundamentally limited by distance.  BLR reverberation mapping is currently the only practical means of probing the SMBH mass function over a large range in redshift and luminosity \citep[][and references therein]{Shen2013TheQuasars}.    

%Typically, $R_{BLR}$ in Equation~\ref{eq:BHmass} is the luminosity weighted radius (LWR, details in Section~\ref{sec:meth:lwr}) measured from the H$\beta$ emission response.  
Arguably the most important result from BLR reverberation mapping is the tight relationship found between the measured lag ($t^\prime$) and the AGN luminosity, which implies a BLR radius – AGN luminosity relationship, $t^\prime\propto R_{BLR} \propto L_{AGN}^{0.5}$ \citep[e.g.,][]{Bentz2009THEH, Bentz2013THENUCLEI}. This provides the key for determining $M_\bullet$ for very large samples, since both the AGN continuum flux and the BEL profile width can be measured from a single spectrum, yielding $L_{AGN}$ (hence $R_{BLR}$) and $\Delta v$, respectively \citep[e.g.,][]{McLure2002MeasuringQuasars, Vestergaard2006DeterminingRelationships, Liu2012ACTIVEACCRETION}.    

% GRAVITY  
In a few nearby and luminous AGN, the BLR has been spatially resolved with near-infrared (NIR) interferometry. Very high resolution interferometry observations of 3C 273, IRAS 09149-6206, and NGC 3783 have been obtained with the GRAVITY instrument using the European Southern Observatory Very Large Telescope Interferometer (VLTI). These data suggest that all 3 AGN have a rotating, disk-like BLR.  From the broad Paschen-$\alpha$ line of the quasar 3C 273, \citet{Sturm2018SpatiallyScale} inferred $R_{BLR}=0.12\pm0.03$\,pc and $M_\bullet = 2.6\pm1.1\times10^8\,M_\odot$.  \citet{Amorim2020The091496206} spatially resolved the broad Br$\gamma$ line in IRAS 09149-6206 and inferred $R_{BLR}=0.075$\,pc and $M_\bullet\sim\times10^8$\,M$_\odot$.  More recently, \citet{Amorim2021TheRegion} used VLTI data to map broad Br$\gamma$ in NGC 3783 and found a BLR with a radial distribution of clouds within $R_{BLR}\approx0.014$\,pc and then in \citet{Amorim2021A3783} reported $M_\bullet=2.54^{+0.90}_{-0.72}\times10^7$\,M$_\odot$.  For all 3 objects, the BLR measurements obtained from GRAVITY are consistent with the $R-L$ relation and reverberation mapping studies of those same AGN.  

\subsection{Reverberation Mapping}
\label{sec:intro:RM}
In the ``point-source reverberation model", clouds respond to fluctuations in the continuum flux on time scales of hours, much smaller than the light crossing time of the whole BLR, which is on the order of days to weeks.  The differences in time for a photon to travel between the accretion disk, a given cloud, and then (as a line photon) to an observer means there is a corresponding time lag between the observed continuum variations and the response of the BELs.  From the perspective of a distant observer, the clouds that respond at the same delay time form parabolic isodelay surfaces, 
\begin{equation}
\label{eq:isodelay}
    t^\prime = \frac{r}{c}(1-\cos\Theta)
\end{equation}
where $r$ is the cloud's radial distance from the continuum source and $\Theta$ is the angle between the observer's line of sight (LOS) and the cloud's position vector. %Therefore, the size of the BLR $R_{BLR}\propto c\Delta t$ 
The observed variation of emission line luminosity with time, $t$, at LOS velocity, $v^{||}$, is given by the convolution of the continuum luminosity, $L_c(t-t^\prime),$ with a transfer function, $\Psi(t^\prime,v^{||})$,      
\begin{equation}
\label{eq:2dtf}
    L(t,\,v^{||}) = \int_{-\infty}^{\infty} \Psi(t^\prime,v^{||})L_c(t-t^\prime)\,dt^\prime
\end{equation}
where $L(t,\,v^{||})$ is the velocity-resolved response of the emission line.  The transfer function is the emission line's response to the continuum at $t=t^\prime$ and encodes information about the BLR's geometry and velocity field \citep{Blandford1982ReverberationQuasars}.  

Typically $R_{BLR}$ is recovered by cross-correlating the observed line light curve with the continuum light curve ($L(t)$ and $L_c(t-t^\prime)$, respectively) to measure the response-weighted time lag, $t^\prime_{RW}$, which is assumed to be equivalent to the luminosity-weighted radius of the BLR; $R_{LW}\approx ct^\prime_{RW}$ \citep{Robinson1990TheVariations,Koratkar1991StructureData,Perez1992TheFunctions,Almeyda2020ModelingFunctions}. The value of $R_{LW}$ inferred is then used in Equation~\ref{eq:BHmass} to find the SMBH mass; i.e., $R_{BLR}=R_{LW}$.  However, due to the complex properties of the BLR, $ct^\prime_{RW}$ is not necessarily an accurate estimate of $R_{BLR}$.  As part of this work, we compare the relation between the $ct^\prime_{RW}$ and $R_{LW}$ values for our BLR models to understand how well the measured lag represents $R_{LW}$. %Since the BLR gas clouds are photoionized by the AGN's UV/optical continuum source, any variation in the continuum corresponds to a radial light-front that propagates outward at speed $c$ from the source to clouds at distance $r$. 
\subsection{Modeling Reverberation Responses \label{sec:intro:mod}}

With sufficient data, the transfer function can be recovered using deconvolution techniques, such as with the \textsc{Memecho} code by \citet{Horne2004ObservationalMapping}.  In practice, however, deconvolving the transfer function accurately from an observed response is not easily done, as it requires very well sampled data over a sufficiently long observing period.  Furthermore, even if the transfer function could be reliably obtained, deciphering the BLR's geometry and velocity parameters would not be a straightforward task, since they would have to be inferred from the shape of the transfer function, which would probably require modeling.  Instead we can use computer simulations to predict $L(t,v^{||})$, given an input AGN light curve, geometry, and velocity field parameters.  The simulated response light curves and BEL profiles for different BLR configurations can then be compared to observed AGN light curves and time-domain spectra.  Numerical reverberation mapping studies of the velocity-resolved response and the transfer function have been presented by  \citet{Welsh1991EchoNuclei,Perez1992TheFunctions,Robinson1995OnNuclei}, and \citet{Peterson2004EchoNuclei}.  More complex reverberation mapping codes, such as the Code for AGN Reverberation and Modeling of Emission Lines (\textsc{Caramel}) by \citet{Pancoast2011GEOMETRICDATA} and, for the AGN torus, the TOrus Reverberation MApping Code (TORMAC) \citep{Almeyda2017DustyAGN}, provide more detailed simulations.

TORMAC, developed by \citet{Almeyda2017ModelingIllumination} and further expanded by \citet{Almeyda2020ModelingFunctions}, simulates the response from the dusty torus for select IR wavelengths given any input light curve.  We have adapted and extended TORMAC for application to the BLR. This new code, the Broad Emission Line MApping Code (BELMAC), generates the velocity-resolved response for various BLR geometrical configurations, gas properties, cloud distributions, and velocity fields, given any input light curve.  The overall goal is to develop BELMAC into a fast and flexible code that can be used to generate large model grids to fit reverberation mapping observations.  However, it is useful to understand how various properties of the BLR affect the reverberation response \citep[e.g.,][]{Welsh1991EchoNuclei,Perez1992TheFunctions,Horne1994EchoSolutions}.  In this paper,  we introduce the basic features of BELMAC and explore the effects of parameters controlling the BLR geometry, kinematics, and cloud properties on the velocity-resolved transfer function. As our main focus here is on the general properties of the BLR we simplify the calculation of the line emission by using hydrogen recombination theory.  A narrow square-wave pulse is used as the as the driving light curve to produce a numerical approximation of the transfer function.

In forthcoming papers, we will describe the next version of BELMAC, which incorporates a grid of photoionization models for more sophisticated emissivity calculations.

BELMAC is outlined in Section~\ref{sec:method} with a description of the geometry setup, velocity field, and the calculation of the cloud emission.  In Section~\ref{sec:results} we present our results for two example models, a spherical BLR with a radiative pressure driven radial outflow and a rotating thin disk. We discuss how our simple reprocessing models can help to understand the behavior of responses and time-domain spectra, the parameters that affect measured time lags, and how BELMAC compares to other reverberation mapping codes in Section~\ref{sec:discuss}. We also discuss how well our results represent various BELs.  Finally, we summarize our conclusions in Section~\ref{sec:concl}. 

\section{Outline of the Code:  BELMAC} 
\label{sec:method}

\begin{figure}[ht]
    \centering
    \includegraphics[width=0.75\textwidth]{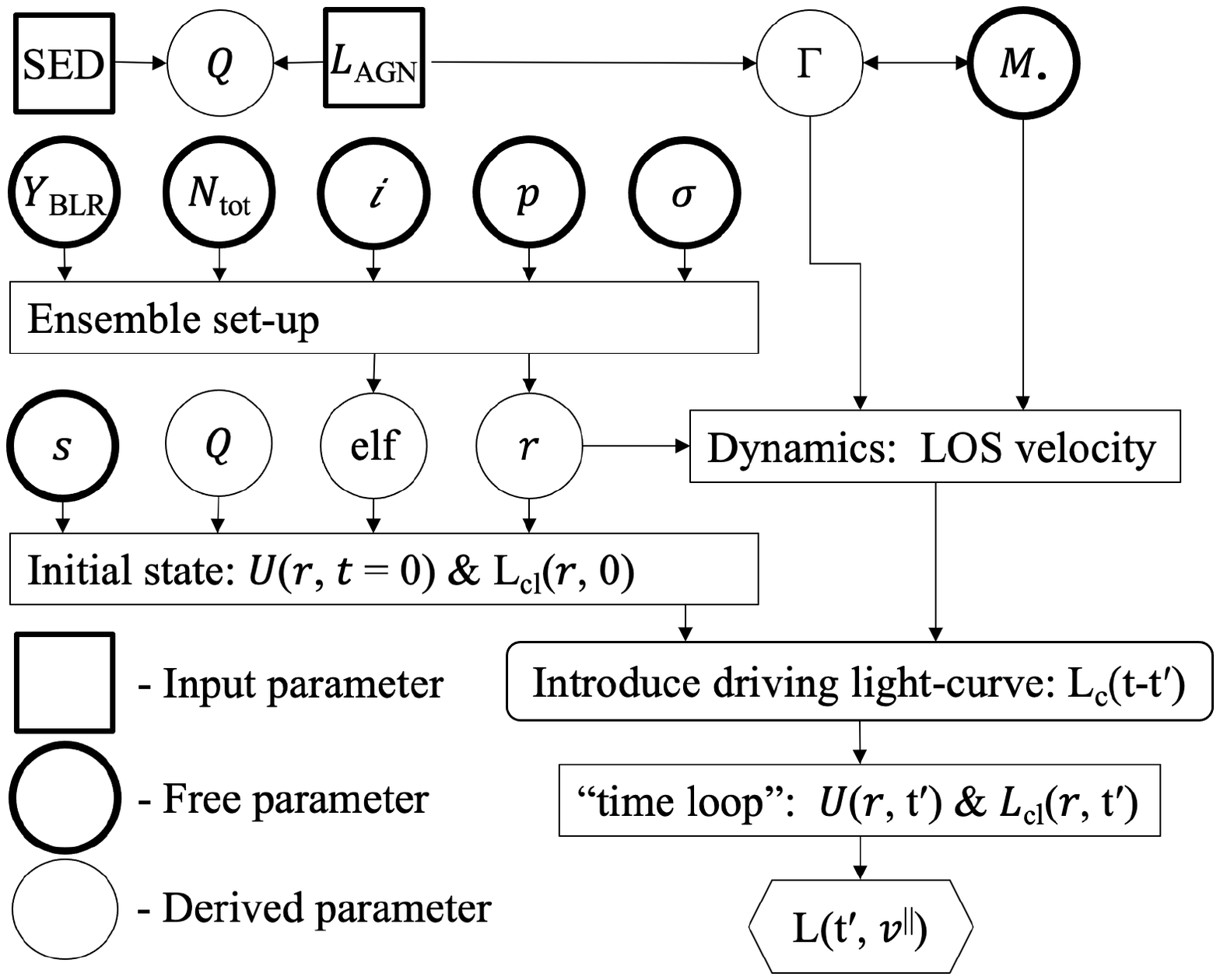}
    \caption{A summary of BELMAC's operation.  The parameters are explained in Section~\ref{sec:method} and listed in Table~\ref{tab:para}.  The parameters in squares are fixed and provided by the user.  The free parameters in thick-lined circles determine the BLR model set-up and parameters in thin-lined circles are calculated within the code.  The final output of the code is the velocity-resolved response, $L(t, v^{||})$.}
    \label{fig:code}
\end{figure}

% overview
The BLR is represented as a 3D ensemble of discrete clouds, that are randomly distributed within a user-defined structure. BELMAC follows the same geometry set-up as TORMAC, but the clouds within the ensemble are dust-free, photoionized gas clouds and have velocities specified by various choices of velocity field.  The general flow of BELMAC is illustrated in Figure~\ref{fig:code} and described in detail in the following subsections.  The user provides the AGN's bolometric luminosity, $L_{AGN}$, spectral energy distribution (SED), and driving light curve.  The geometry, velocity field, and distribution of the ensemble of BLR clouds are described by $Y_{BLR},~\sigma,~i,~C_f,~s,~p$, and $M_\bullet$. The cloud positions are randomly generated and each cloud's LOS velocity is calculated based on the type of velocity field specified, either Keplerian, radial flow, or turbulent.  In all, there are 10 descriptive parameters for each distinctive BLR model.    

BELMAC can use any input light curve, but in order to obtain a numerical approximation of the transfer function, the input light curve is a single square-wave pulse to represent a $\delta$-function. We refer to the approximate transfer function as the response function \citep{Almeyda2017ModelingIllumination}.  For the duration of the pulse, $\sim4$ days, the ionizing luminosity increases by a factor of 2.  The light-front corresponding to the light pulse propagates outward from the center of the system with time and the total luminosity of the BLR is computed by integrating over the cloud ensemble at each observer time step, taking into account light travel delays.   

\subsection{BLR Geometry and Cloud Structure}
\label{sec:meth:geo}

\begin{figure}[ht]
    \centering
    \includegraphics[width=0.75\textwidth]{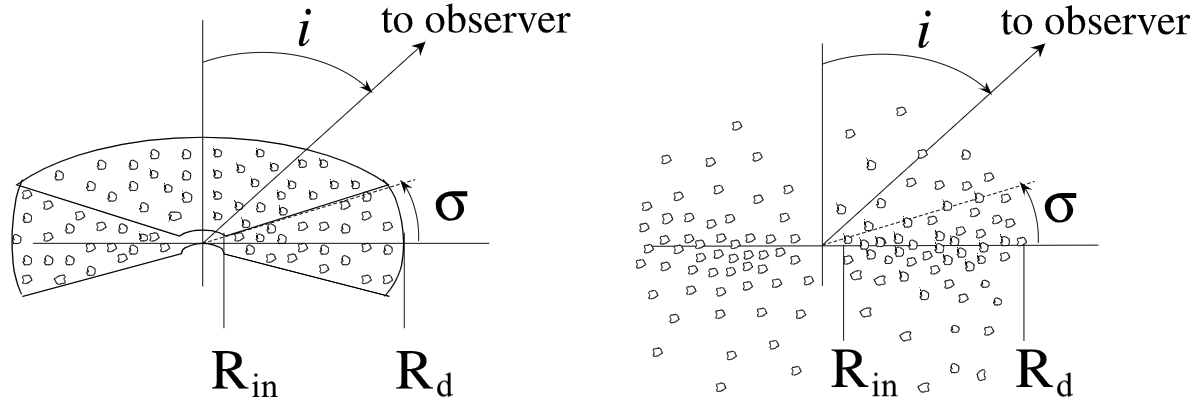}
    \caption{Geometrical parameters used for clouds arranged in a flared disk in BELMAC. $R_{in}$ and $R_{d}$ are the inner and outer radii, $\sigma$ is half-angle width of the disk, and $i$ is the inclination of the disk axis to the observer's line of sight. On the left is a sharp edged BLR configuration and a `fuzzy' (Gaussian distribution in $\beta$) boundary is shown on the right.  Adapted from \citet{Nenkova2008AGNCLUMPINESS}.}
    \label{fig:ensemble}
\end{figure}

In common with many other studies \citep[e.g.,][]{Armijos-Abendano2022Broad-lineSample,Marconi2008TheGalaxies,Robinson1990TheVariations,Davidson1979TheObjects}, we assume that the BLR gas is distributed in many small clouds. The ensemble of discrete, gas clouds are randomly distributed within an overall geometry of either a disk or a sphere.  Cloud positions are defined by the spherical polar coordinates $r,\,\theta,\,\phi$, where $r$ is the radial distance from the continuum source at the origin, $\theta$ is the polar angle, and $\phi$ is the azimuthal angle. In the models considered here, the BLR is assumed to have `sharp' boundaries with the clouds uniformly distributed in elevation above the equatorial plane, $90^o-\theta$, and in $\phi$.  However, `fuzzy' edges may also be arranged by drawing values of the elevation angle from a Gaussian distribution \citep[Figure~\ref{fig:ensemble};][]{Almeyda2017ModelingIllumination}.  The disk's angular width is set by the half-angle $\sigma$, where $\sigma=90^o$ is a spherical ensemble (Figure~\ref{fig:ensemble}).  The inclination, $i$, is the angle between the polar axis and the observer's LOS, therefore, for $i=0^o$ the disk is viewed face-on and for $i=90^o$ it is viewed edge-on.  

The outer radius of the BLR is set by the dust sublimation radius, $R_d$,  
\begin{equation}
\label{eq:Rd}
    R_{d}=0.4 \left( \frac{L_{AGN}}{10^{45}\,\mathrm{erg\,s^{-1}}}\right) ^{1/2} \left( \frac{1500\,K}{T_{sub}}\right) ^{2.6} \mathrm{pc} 
\end{equation}
where $T_{sub}$ is the dust sublimation temperature \citep{Nenkova2008AGNMEDIA}.  Here we take representative values,  $L_{AGN}=10^{45}$\,erg\,s$^{-1}$ and T$_{sub}=1500$\,K, then $R_d\approx10^{18}$\,cm (0.4\,pc).  For simplicity, we will treat the dust sublimation radius as a sharp boundary between the BLR and torus, rather than as a transition region, which would be more physically realistic \citep[][]{Baskin2018DustNuclei}.  The inner radius of the BLR, $R_{in}$, is set by the free parameter $Y_{BLR} = R_d/R_{in}$.  In all of the models presented here, we will use $Y_{BLR}=20$;  similar to scaled sizes used in previous modeling \citep[e.g.][]{Du2015SUPERMASSIVEACCRETION,Netzer2020TestingMapping}, giving $R_{in}\approx6\times10^{16}$\,cm ($0.02$\,pc), which corresponds to a light-crossing time $\sim24$\,days.  This is consistent with observed BLR time-delays that range from days to $\sim$a month \citep[e.g.,][]{Fausnaugh2017ReverberationGalaxies} and the $R-L$ relation \citep{Bentz2013THENUCLEI}.      

The radial cloud distribution is given by a power-law with index $p$, such that the number of clouds between $r$ and $r+dr$ is, 

\begin{equation}
\label{eq:cloud distribution}
    N(r)dr = A_o \left( \frac{r}{R_d} \right)^pdr 
\end{equation}

where $A_o$ is a normalization constant that is determined by the total number of clouds, $N_{tot}=\int_{R_{in}}^{R_d}N(r)dr$ \citep{Almeyda2017ModelingIllumination}. 
The BLR clouds are assumed to be spherical, with radii, $R_{cl}(r)$, specified by the integrated covering fraction of the BLR, 
 
\begin{equation}
\label{eq:CF}
    C_f = \int_{R_{in}}^{R_d} \frac{N(r)}{4\pi r^2}A_{cl}(r)\sin\sigma dr 
\end{equation}

where the cloud's cross-sectional area is $A_{cl}(r)=\pi R^2_{cl}(r)$.  We assume the clouds are composed of pure hydrogen and have a constant mass, $m_{cl}(r)=m_{cl}$.  For a pure hydrogen gas cloud, the ionized portion of the cloud has an electron density equal to the hydrogen density, which is the cloud's gas density $n(r)$. The gas density is a power-law function of the cloud's radial distance from the continuum, $n(r)=n_o(r/R_d)^s$, where the values of the gas density at the outer radius of the BLR, $n(R_d)=n_o$, and the power-law index, $s$, are free parameters. Therefore, the cloud radius $R_{cl}(r) \propto r^{-s/3}$ and the cloud column density,

\begin{equation}
\label{eq:density}
    N_{cl}(r) \propto R_{cl}(r)n(r) \propto r^{2s/3}
\end{equation}

will vary with distance from the ionizing source.  Realistically, the mass of the clouds could change with distance depending on how the clouds are confined, such as by radiation pressure or magnetic fields \citep{Netzer2008IonizedNuclei}.     
Table~\ref{tab:para} summarizes the BELMAC geometry parameters with their respective descriptions, symbol, and values we will adopt as ``standard". 

\begin{table}[ht]
\caption{BLR geometry and property parameters used to set-up for the ensemble and their standard model value.}
\begin{tabular}{l|c|l}
\label{tab:para}
    Parameter Description & Symbol & Standard Values \\
    \hline
    Outer radius & $R_d$ & $1.23\times10^{18}$\,cm  \\
    Size scaled to $R_d$ & $Y_{BLR}$ & 20  \\
    Angular width & $\sigma$ & $10^o-90^o$  \\
    Cloud distribution power-law index & $p$ & 0, 1, \& 2 \\
    Total number of clouds & $N_{tot}$ & 100,000 \\
    Inclination to observer & $i$ & $0^o-90^o$  \\
    Covering fraction &  $C_f$ & 0.3 \\
    Cloud sizes &  $R_{cl}(r)$ & $10^{12}-10^{14}$\,cm \\
    Gas density in a cloud at $R_d$ & $n(R_d)$ & $10^9$\,cm$^{-3}$ \\
    Total ionizing photon luminosity \footnote{Determined from a user specified SED and $L_{AGN}$} & $Q_H(R_d,t^\prime=0)$ & $2.6\times10^{55}$\,photons\,s$^{-1}$ \\
    Power-law index for gas density  & $s$ & 0, -1, \& -2\\
    Emission line fraction \footnote{Not a free parameter, but can be turned on or off} & $elf$ & $0-1$\\
\end{tabular}
\end{table}

\subsection{Cloud Emission}
\label{sec:meth:emission} 

To approximate the H$\alpha$ luminosity of a cloud, we use hydrogen recombination theory \citep{Osterbrock1986Emission-LineQSOs}.  For spherical clouds, $d_c(r)$ is the average path length through the cloud, which is proportional to $R_{cl}(r)$.  The surface of a cloud facing the source is exposed to the ionizing flux and is entirely ionized.  Assuming ionization equilibrium, the clouds are fully ionized within the Str\"omgren depth, 
\begin{equation}
    d_s(r) = \frac{cU(r,t^\prime)}{n(r)\alpha_B} = N_s(r)/n(r)
\end{equation}
where $\alpha_B$ is the recombination coefficient for hydrogen ($\alpha_B\approx 2.6\times10^{-13}$\,cm$^3$\,s$^{-1}$; \citet{Osterbrock1986Emission-LineQSOs}) and $N_s(r)$ is the Str\"omgren column density.  The ionization state of a given cloud is described by the ionization parameter, the ratio of the local ionizing photon flux, $\Phi(r,t^\prime)$, to the cloud's gas density, 

\begin{equation}
\label{eq:U}
    U(r,t^\prime) = \frac{\Phi(r,t^\prime)}{cn(r)} = \frac{Q_H(t^\prime)}{4\pi r^2cn(r)} \propto r^{-s-2}
\end{equation}
where $Q_H(t^\prime)$ is the total ionizing photon luminosity.  Since the continuum luminosity naturally undergoes temporal variability, $Q_H(t^\prime)$ and, hence, $U(r,t^\prime)$ fluctuate with time.  $Q_H(t^\prime=0)$ is obtained from any selected AGN's SED and $L_{AGN}$, which sets $U(R_d,0)$.  To estimate $Q_H(0)$, for the models presented in this paper, we use the compiled SED created by \citet{Jin2012AProperties} who constructed an average SED using 17 type 1 AGN that have an average Eddington ratio $\Gamma=0.07$ and $M_\bullet=10^8$\,M$_{\odot}$.   

If $d_s\geq d_c$, the cloud is fully ionized, or ``matter-bounded'', and has reached its maximum ability to emit hydrogen recombination line radiation. Otherwise a cloud is partially ionized and ``radiation-bounded."  The cloud luminosity for radiation and matter-bounded clouds is, respectively,

\begin{equation}
    \label{eq:lum}
    L_{cl} =\left\{ 
\begin{array}{ll}
      n(r)^2\alpha_{H\alpha}^{eff}h\nu_{H\alpha} \pi R^2_{cl}(r)d_s(r) & \mathrm{for}~ d_s < d_c  \\ \\
      n(r)^2\alpha_{H\alpha}^{eff}h\nu_{H\alpha} \pi R^2_{cl}(r)d_c(r) & \mathrm{for}~ d_s \geq d_c 
\end{array} 
\right.
\end{equation}

where $\alpha_{H\alpha}^{eff}$ is the effective recombination coefficient for H$\alpha$ ($\alpha_{H\alpha}^{eff}\approx 1.1\times10^{-13}$\,cm$^3$\,s$^{-1}$). In the matter-bounded case for constant mass clouds, the cloud luminosity $L_{cl}(r)\propto n(r) \propto r^s$, but is independent of $U(r,t^\prime)$ and therefore is not time-dependent.  In the radiation-bounded case, the cloud luminosity does depend on $U(r,t^\prime)$ and thus $L_{cl}(r,t^\prime)\propto n(r)A_{cl}(r)U(r,t^\prime) \propto r^{-2(1+s/3)}Q_H(t^\prime)$.  The cloud luminosities are summed at each time intervals in the observer's frame to calculate the total BLR luminosity as a function of time.  

\begin{figure}[ht]
    \centering
    \includegraphics[trim=0 0 290 0,clip,width=0.7\columnwidth]{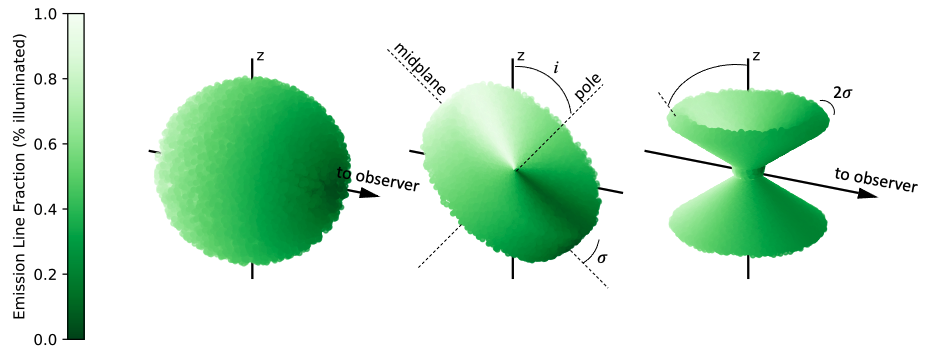}
    \caption{Two geometrical BLR configurations explored in this paper, a sphere and  thin disk ($i=45^o$, $\sigma=15^o$).  The color represents the $elf$ parameter.  The dark green means the observer views only the non-illuminated face of the cloud.  The white means the observer sees only the fully illuminated side of the cloud. Intermediate shades indicate the fraction of the illuminated side see by the observer.}
    \label{fig:elf}
\end{figure}

At the high densities of the BLR, some emission lines, such as the Balmer lines, may become optically thick~\citep[][]{Osterbrock1986Emission-LineQSOs}. It is also possible that dust grains may survive in the cooler, largely neutral gas beyond the ionization front \citep[e.g.,][]{Baskin2018DustNuclei}, causing extinction of line radiation produced in the ionized zone. As a result, the line emission of a cloud may vary strongly between the illuminated and non-illuminated faces. To account for this effect, BELMAC includes an approximate treatment of anisotropic cloud emission (ACE), which is dependent on a cloud's location with respect to the continuum source and the observer's line of sight.  Each cloud's luminosity is multiplied by its emission line fraction ($elf$), 

\begin{equation}
    \label{eq:elf}
    elf =\left\{ 
\begin{array}{ll}
      \frac{1}{\pi}\arccos(\sin\theta\cos\phi) & ~~\mathrm{for}~ d_s < d_c  \\
      1 & ~~\mathrm{if}~ d_s \geq d_c 
\end{array} 
\right.
\end{equation}

which is the fraction of the illuminated-side an observer sees.  However, if a cloud is matter-bounded ($d_s \geq d_c$) then we assume that it has isotropic cloud emision (ICE) and set $elf=1$.  Figure \ref{fig:elf} shows from the observer's perspective how the $elf$ changes the cloud's luminosity with position.     

We summarize the parameters governing a cloud's emission and internal properties, along with their typical values in Table~\ref{tab:para}. 

% emissivity
\subsection{The Analytical Transfer Function}
\label{sec:meth:tf}

To understand how the basic features of the response function are influenced by the size of the BLR, the gas density, and radial cloud distribution, we discuss the analytical transfer function for a spherical shell.  The radial cloud distribution and internal cloud properties determine the volume emissivity, $\varepsilon_V(r) \propto L_{cl}N(r)/r^2$, which will affect the shapes of the line profile and transfer function.  We adopt \citet{Perez1992TheFunctions}'s assumption that the cloud's emission efficiency behaves as a power-law with distance and approximate the volume emissivity as,     

\begin{equation}
\label{eq:emiss}
    \varepsilon_V(r) = \varepsilon_o (r/R_{in})^\eta
\end{equation}

where $\varepsilon_o=\varepsilon_V(R_{in})$ and the emissivity index $\eta$ relates to our $p$ and $s$ indices (from Equations~\ref{eq:cloud distribution} and~\ref{eq:density}, respectively) as,

\begin{equation}
    \label{eq:eta:tf}
    \eta =\left\{ 
\begin{array}{ll}
      p-4-2s/3 & \mathrm{for}~ d_s < d_c  \\
      p-2+s & \mathrm{for}~ d_s \geq d_c \\
\end{array} 
\right.
\end{equation}

for radiation and matter-bounded clouds, respectively. Recalling Equations~\ref{eq:isodelay} and ~\ref{eq:2dtf}, the analytical transfer function is given by,

\begin{equation}
\label{eq:ana:tf}
    \Psi(t^\prime) = \int_V \varepsilon_V(r) \delta(t-t^\prime) dV
\end{equation}

where $V$ is the volume of the BLR.  Solving Equation~\ref{eq:ana:tf} for a spherical shell BLR model we arrive at, 

\begin{equation}
\label{eq:ana:tf:sphere1}
   % \begin{align}
    \Psi(\tau) \propto 
    \begin{cases}
              Y_{BLR}^{\eta+2}-1 & \mathrm{for}~ \tau \leq 1/Y_{BLR}~~\&~\eta \neq -2 \\
              Y_{BLR}^{\eta+2}(1-\tau^{\eta+2}) & \mathrm{for}~ \tau > 1/Y_{BLR}~~\&~\eta \neq -2\\
        \end{cases}  
    %\end{align}
\end{equation}
and
\begin{equation}
\label{eq:ana:tf:sphere2}
    %\begin{align}
    \Psi(\tau) \propto 
        \begin{cases}
              \ln{Y_{BLR}} & \mathrm{for}~ \tau \leq 1/Y_{BLR}~~\&~\eta = -2 \\
              \ln{1/\tau} & \mathrm{for}~ \tau > 1/Y_{BLR}~~\&~\eta = -2 
        \end{cases}  
    %\end{align}
\end{equation}

where $\tau = ct^\prime/2R_d$ is the delay in units of the light crossing time of the BLR, such that the line response is complete when $\tau = 1$ \citep{Robinson1990TheVariations,Perez1992TheFunctions,Almeyda2017ModelingIllumination}.  In terms of $\tau$, the duration of the square-wave pulse used in our models is 0.004.  

% LWR and RWD
\subsection{Response Weighted Delay and Luminosity Weighted Radius}
  
\begin{figure}[ht]
    \centering
    \includegraphics[width=0.6\columnwidth]{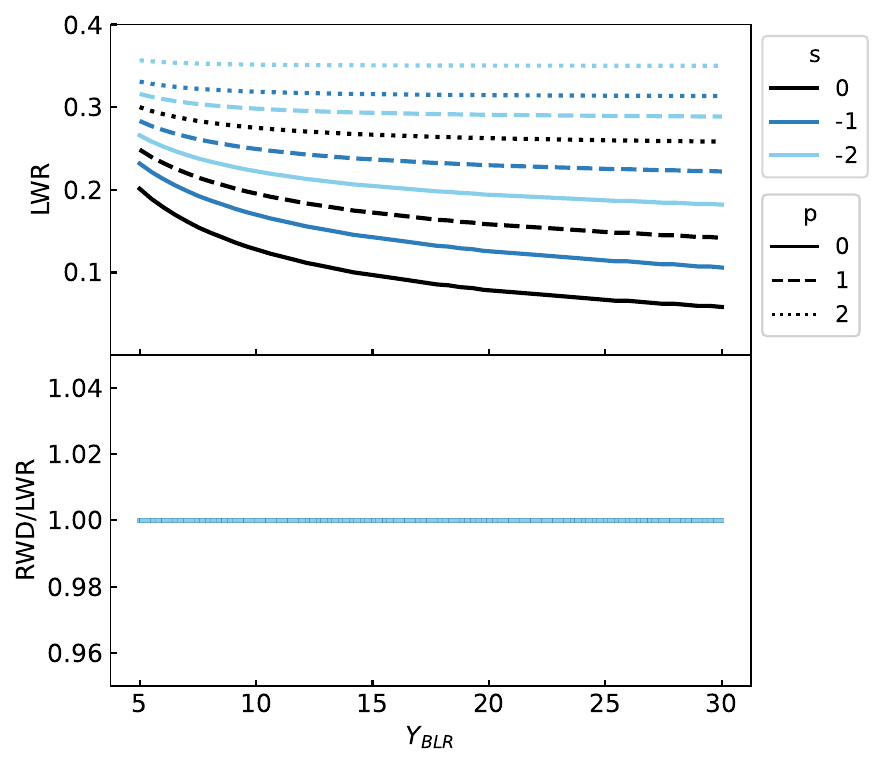}
    \caption{The analytical LWR (top row) and ratio RWD/LWR (bottom row) for a spherical shell for varying combinations of $s$ and $p$ parameters with $Y_{BLR}$. }
    \label{fig:ana:LWR}
\end{figure}

As mentioned in Section~\ref{sec:intro:RM}, the response-weighted time delay determined from reverberation mapping is considered a reasonable estimate of the radial extent of the BLR, $ct^\prime_{RW}\approx R_{LW} = R_{BLR}$.  However, it is unclear how accurate this assumption is due to the uncertain and complex cloud properties of the BLR. \citet{Robinson1990TheVariations} defined the dimensionless, response-weighted delay (RWD) as,       
 
\begin{equation}
\label{eq:rwd}
    \mathrm{RWD}=\frac{\displaystyle\int_0^1\tau\Psi(\tau)d\tau}{\displaystyle\int_0^1\Psi(\tau)d\tau}
\end{equation}  
to characterize the delay associated with the transfer function. This is related to the cross-correlation lag as $t^\prime_{RW}=2R_{d}$RWD$/c$ in physical units. For a spherical shell containing isotropically emitting clouds, the RWD is exactly the dimensionless luminosity-weighted radius (LWR),             
\begin{equation}
\label{eq:lwr}
    \mathrm{LWR}=\frac{\displaystyle\int_Vr\varepsilon_VdV}{\displaystyle2R_{d}\int_V\varepsilon_VdV}
\end{equation}
and $R_{BLR}=2R_{d}$LWR \citep{Robinson1990TheVariations,Almeyda2020ModelingFunctions}.  The analytical LWR and RWD/LWR for spherical shell BLRs with varying $Y_{BLR},~s$, and $p$ are shown in Figure~\ref{fig:ana:LWR}.  We will present and compare the LWR and RWD of various BLR models in Section~\ref{sec:results}.

\subsection{BLR Cloud Dynamics}
\label{sec:meth:vel}

In order to compute the velocity-resolved response, we include three types of velocity fields; rotational, radial, and random turbulence.  The broadest line widths observed are $\lesssim 0.1c$ \citep[e.g.,][]{Assef2011BLACKLINES,Fausnaugh2017ReverberationGalaxies}, hence the dynamical time scale of the BLR is much greater than ($\gtrsim 10\times$) its light-crossing time.  Therefore, we assume the clouds are static as they do not significantly move along their trajectories over timescales typically of interest for reverberation mapping.  

BELMAC is capable of combining radial, rotational, and random motions for a multi-component velocity field, however, for the purposes of this paper, we will present and discuss each type of motion separately.  Furthermore, we will not explore the random, turbulent motion in the models presented here.  

\subsubsection{Rotational Velocity Field}
\label{sec:vel:rot}

 In the rotating disk BLR, the clouds are assumed to follow circular, Keplerian motion,  $v_{Kep}(r)=\left(GM_\bullet/r\right)^{1/2}$.  The rotational velocity is scaled to the velocity at the inner radius,
 \begin{equation}
 \label{eq:vel:rot}
 v_{rot}(r) = v_{Kep}(R_{in})(r/R_{in})^{-1/2}
 \end{equation}
 for a cloud at radius, $r$.  The cloud orbits are randomly inclined relative to the disk's mid-plane and are constrained by the disk angular width, $\sigma$, which requires $90^o-\sigma \leq |\theta| \leq 90^o+\sigma$. Given a cloud with coordinates $r$, $\theta$, and $\phi$ and a BLR inclination $i$, the velocity component along the LOS velocity to an observer is,
\begin{equation}
\label{eq:vel:rot:los}
    v^{||}_{Kep} = -v_{rot}(r)(\cos\theta \sin\phi \cos i - \sin\theta \sin\phi \sin i).
\end{equation} 

If we combine Equations~\ref{eq:isodelay} and \ref{eq:vel:rot:los} for an infinitesimally thin disk ($\sigma\to 0, \theta=90^o$),  
 
\begin{equation}
\label{eq:ellipse}
    \left(\frac{ \displaystyle v_{Kep}^{||}}{\displaystyle v_{rot}\sin i} \right)^2+\frac{\displaystyle (t^\prime-(r/c))^2}{\displaystyle (r/c\,\sin i)^2} = 1
\end{equation}

we find the response for a given orbital radius in the projected velocity and time delay space forms an ellipse.  The ellipses are centered at $(t^\prime=r/c,~v^{||}_{Kep}=0)$, with semi-axes $\pm v_{rot}\sin i$ in velocity and $r/c\,\sin i$ in time delay \citep{Perez1992TheFunctions,Welsh1991EchoNuclei}. 

\subsubsection{Radial Velocity Field}
\label{sec:vel:rad}

The radial motion is dependent on the primary outward driving mechanism, assumed here to be radiation pressure, and the inward pull of gravity.  We further assume that the BLR intercloud medium is sufficiently tenuous that we can ignore the forces due to drag and the pressure gradient between the clouds.  We follow \citet{Marconi2008TheGalaxies}'s and \citet{Netzer2010THENUCLEI}'s prescription for clouds accelerated by radiation pressure due to Thomson scattering and absorption of ionizing radiation emanating from the accretion disk. 

\begin{figure}[ht]
    \centering
    \includegraphics[width=.8\textwidth]{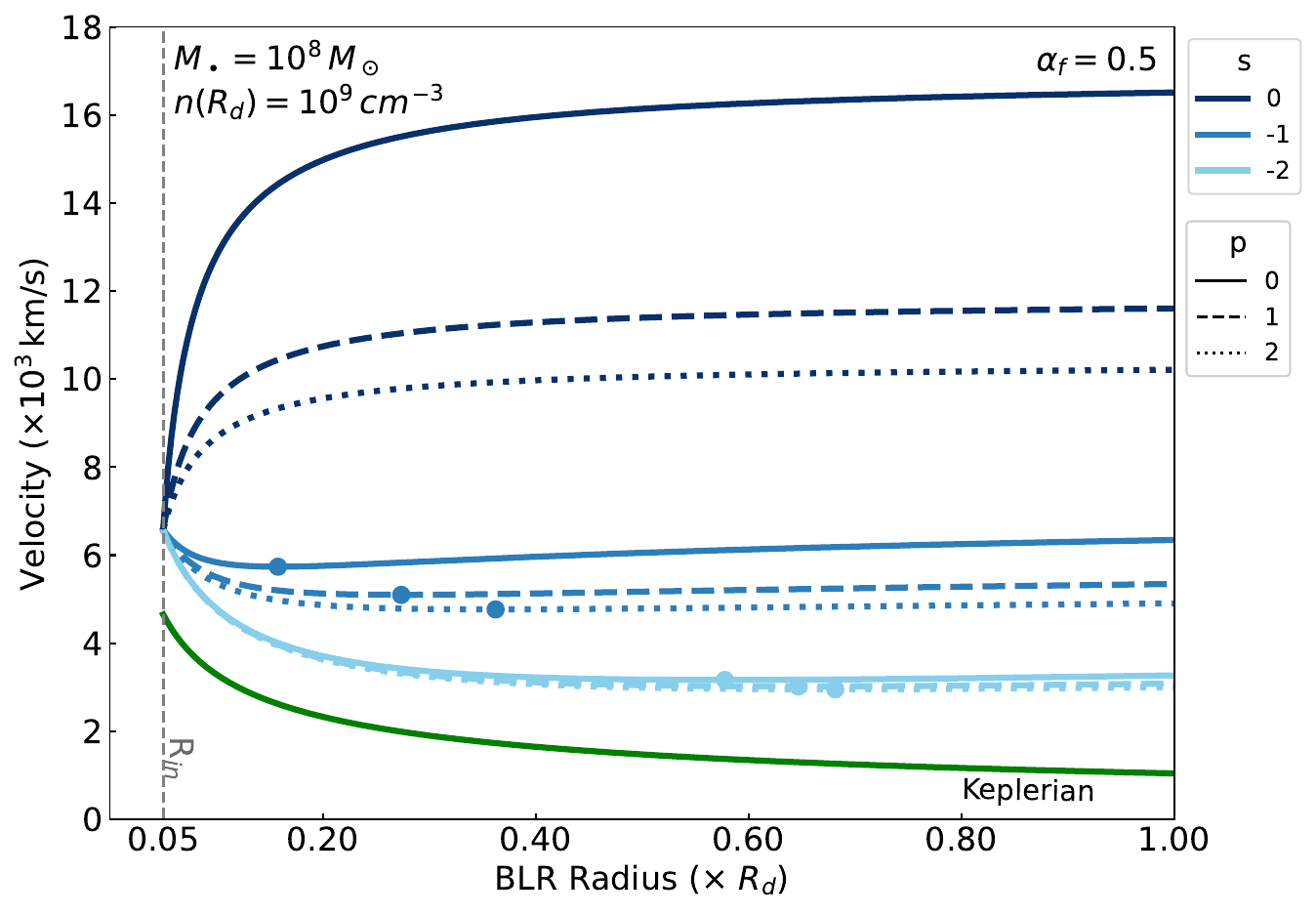}
    \caption{Rotational and radial velocities with respect to distance scaled to $R_d$.  The escape velocity at $R_{in}$ (dashed gray line) is 6,578\,km\,s$^{-1}$.  The blue curves show the radial velocity given by Equation~\ref{eq:vel:rad}.  The lighter the shade, the lower the $s$ index.  The solid, dashed, and dotted lines are $p=0,~1$, and 2, respectively.  The blue circles on the $s=-1$ and $-2$, $p=0,~1,$ and 2 curves mark the distance at which radiation pressure overcomes gravity and the clouds start accelerating outwards.  Each of these points corresponds to the critical $N_{col}=5.26\times10^{22}$\,cm$^{-2}$.  When $s=0$, the outward, radiation pressure force is greater than gravity at all distances.  The green curve is the rotational velocity field, $v_{Kep}$, described by Equation~\ref{eq:vel:rot}, which does not depends on the $s$ and $p$ indices.  
    }
    \label{fig:vel}
\end{figure}

The equation of motion for a cloud of pure hydrogen gas at a distance $r$ from the radiation source, with cross-sectional area $A_{cl}(r)$ and mass $m_{cl}$ is, 

 \begin{equation}
  \label{eq:acc:total}
    a(r) = \frac{A_{cl}(r)\alpha_fL_{AGN}}{4\pi r^2m_{cl}c} - \frac{GM_\bullet}{r^2}
 \end{equation}
  
where $\alpha_f \simeq \int_{\nu_H}^\infty L_\nu d\nu / L_{AGN}$, the fraction of the AGN's bolometric luminosity that is absorbed by the cloud.  Although the absorption fraction will vary with $U(r,t^\prime)$ and $N_{col}(r)$, in this paper we use the constant value $\alpha_f=0.5$, following \citet{Netzer2010THENUCLEI}, throughout the BLR. 
 Putting Equation~\ref{eq:acc:total} in terms of $N_{col}$ and recognizing that the Eddington luminosity is $L_{Edd}=4\pi GM_\bullet m_Hc/\sigma_T$, Equation~\ref{eq:acc:total} becomes, 
 
 \begin{equation}
 \label{eq:eom}
     \frac{dv_{rad}}{dt} = \frac{GM_\bullet}{r^2} \left(\frac{\alpha_fL_{AGN}}{\sigma_TN_cL_{Edd}} -1 \right) = \frac{GM_\bullet}{r^2} \left(\Gamma F_M -1 \right)
 \end{equation}
after simplifying by introducing the Eddington ratio, $\Gamma=L_{AGN}/L_{Edd}$, and the force multiplier, $F_M=\alpha_f/\sigma_TN_c$, where $\sigma_T$ is the Thomson cross-section.  It can now be easily seen that for $\Gamma F_M>1$ the cloud is accelerated outwards and accelerated inwards when $\Gamma F_M<1$ \citep{Marconi2008TheGalaxies,Netzer2008IonizedNuclei,Netzer2010THENUCLEI}. The radii at which $\Gamma F_M=1$ are when $N_{col}=5.26\times10^{22}$\,cm$^{2}$ are are indicated in Figure~\ref{fig:vel}; for the $s=-1$ models at $r=2.55\times10^{17}$\,cm, $7.12\times10^{17}$\,cm, and $5.84\times10^{17}$\,cm, and for $s=-2$, at $r=8.15\times10^{17}$\,cm, $9.12\times10^{17}$\,cm, and $9.61\times10^{17}$\,cm for $p=0,~1,$ and 2, respectively.
 
If the BLR is created by a wind from the accretion disk \citep[e.g.,][]{Elvis2000AQuasars}, the gas clouds at $R_{in}$ would have a velocity comparable to the escape velocity from the disk.  Therefore, we assume the radial velocity at $R_{in}$ to be the local escape velocity.  Integrating Equation~\ref{eq:eom} we obtain,

\begin{eqnarray}
\label{eq:vel:rad}
    v_{rad}(r)^2 = \frac{2GM_\bullet}{r}\left[\frac{\Gamma F_M}{2s/3+1}\left(\left(\frac{r}{R_{in}}\right)^{2s/3+1}-1\right)+1
    \right]
\end{eqnarray}

the radial velocity field.  Figure~\ref{fig:vel} shows $v_{Kep}(r)$ and $v_{rad}(r)$, Equations~\ref{eq:vel:rot} and \ref{eq:vel:rad}, respectively, for several values of the gas density and cloud radial distribution power-law indices.               

The component of the radial velocity along the observer's LOS is given by, 
\begin{equation}
\label{eq:vel:rad:los}
    v^{||}_{rad} = -v_{rad}(r)(\cos\theta \cos i + \sin\theta \cos\phi \sin i).
\end{equation}
Combining Equations~\ref{eq:vel:rad:los} and~\ref{eq:isodelay} the delay-LOS velocity relation for a thin, spherical shell with radius $r$ is a straight line in velocity and time delay space,  

\begin{equation}
    \tau = \frac{r}{c}\left(1+ \frac{v^{||}_{rad}}{v_{rad}}\right)
\end{equation}
 
where the slope, $\frac{r}{c\,v_{rad}}$, is positive for an outflow and negative for an inflow \citep{Welsh1991EchoNuclei,Perez1992TheFunctions}.  

\section{Results} \label{sec:results}

% Standard Model Table
\begin{table}[ht]
\label{tab:SM}
\caption{Standard model parameters for a spherical and disk-like BLR.  Fixed values for all geometries are $C_f$ = 0.3,  $L_{AGN}=10^{45}$\,erg\,s$^{-1}$, $\Gamma=0.07$, $\alpha_f=0.5$, $Y_{BLR}=20$, $R_d\approx10^{18}$\,cm, and a representative sub-sample of 100,000 clouds.}
\begin{tabular}{c|c|c}
    Parameter Description & \multicolumn{2}{c}{Standard Values} \\
    \hline
    Parameter & Sphere & Disk \\
    \hline
    Bulk motion & radial & Keplerian \\
    Inner radius velocity $v(R_{in})$ & $6,578$\,km\,s$^{-1}$ & $4,650$\,km\,s$^{-1}$ \\
    %Max velocity ($v_{max}$) & 5,000\,km\,s$^{-1}$ & 5,000\,km\,s$^{-1}$ \\
    Half-angle width ($\sigma$) & $90^o$ & $15^o$  \\
    Inclination ($i$) & - & $0^o - 90^o$ \\
    Cloud initial gas density ($n(R_d)$) & $10^9$cm$^{-3}$ & $10^9$cm$^{-3}$ \\
    Cloud distribution index ($p$) & 0, 1, 2 & 0, 1, 2 \\
    Gas density index ($s$) & 0, -1, -2 & 0, -1, -2 \\
    Total ionizing  $(Q_H(r,\tau=0))$ \footnote{Calculated from the SED model by \citet{Jin2012AProperties} for $M_\bullet=10^8$\,M$_\odot$ and $\Gamma=0.1$.  See their Figure 14, panel 4-C.}  & $2.6\times10^{55}$\,erg\,s$^{-1}$ & $2.6\times10^{55}$\,erg\,s$^{-1}$ \\
    Ionization parameter ($U(R_d,\tau=0))$) & $4.5\times10^{-2}$ & $4.5\times10^{-2}$ \\
    \hline
\end{tabular}
\end{table}

Here we present the velocity-resolved reverberation response functions (2DRFs) of, firstly, a spherical shell BLR with a radiation pressure driven outflow and, secondly, a rotating disk BLR, for their respective parameters listed in Table~\ref{tab:SM}.  We also present the velocity-integrated response functions (1DRFs), time-averaged line profiles, and root-mean-square (RMS) line profiles.     

To isolate the response amplitude, we subtract the initial BLR state (prior to the onset of the continuum pulse) from the response function at each $\tau$. The 1DRF is normalized as $L(\tau)_{norm}=\frac{L(\tau,\,v^{||})-L_o}{L_{max}-L_o}$, where $L_o$ is the BLR luminosity in its initial state and $L_{max}$ is the peak luminosity of the BLR's response.  We will present results from BLR models with both isotropic cloud emission (ICE) and anisotropic cloud emission (ACE).  The BLR models presented here include $N_{tot}=10^5$ clouds, but as the BLR probably contains $\sim 10^6$ clouds \citep{Arav1998Are4151,Dietrich1999Structure273}, these can be considered a representative sub-sample.        

\subsection{Spherical BLR with Radial Outflow}
\label{sec:res:sphere}

\begin{figure}
%\begin{interactive}{animation}{Figures/SphericalOutflow_with_Isodelay.mp4}
\includegraphics[width=\textwidth]{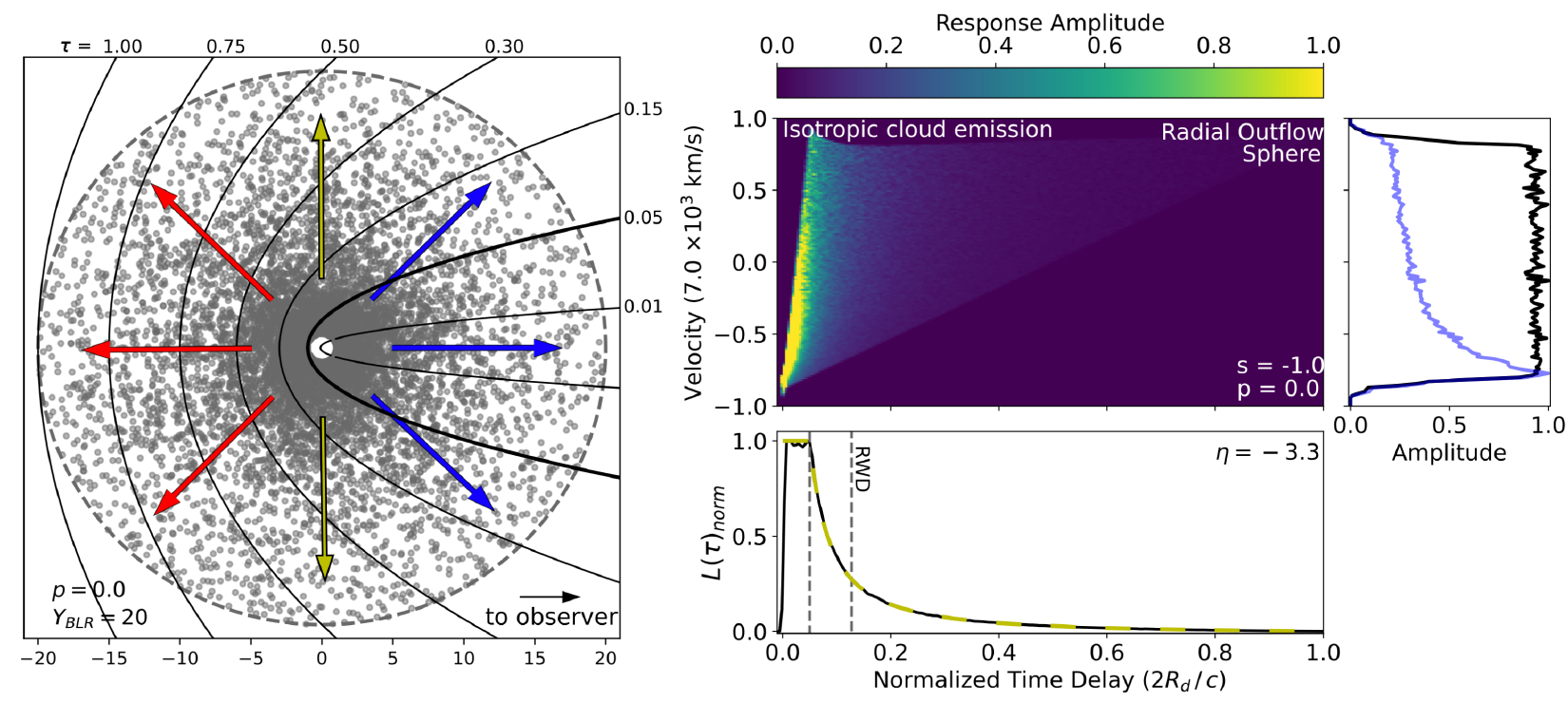}
%\end{interactive}
\caption{\textit{Left}: Cross-section of a spherical BLR for $Y_{BLR}=20$ with isodelay surfaces over plotted for several values of the normalized time delay, $\tau$. The arrows represent the velocity field, with the colors blue, red, and yellow indicate a blueshift, redshift, and no shift, respectively.  An animation of this figure can be viewed in the online version of this paper.  In the animation, the isoddelay surfaces appear with $\tau$ from 0 to 1, which is denoted in the upper left-hand corner. In the static figure, all the isodelay surfaces are shown and each isodelay curve is labeled with it's corresponding value $\tau$ along the border.  The bold isodelay curve corresponds to the light-crossing time of $R_{in}$, $\tau=0.05$.  \textit{Right}: Response function for a spherical BLR with a radial outflow, for a gas density index $s=-1$ and cloud distribution index $p=0$.  All of the clouds in this BLR model are radiation-bounded.  The color-scale represents the normalized response amplitude.  The lower panel shows the velocity averaged, 1DRF in black, with the corresponding analytical model, calculated using Equation~\ref{eq:ana:tf} with $\eta=-10/3$, in dashed yellow.  In the animated version, the 1DRF is plotted in-sync with the isodelay surfaces in the left figure. The dashed lines in the 2D and 1DRFs indicate the light-crossing time of $R_{in}$, $\tau = 1/Y_{BLR}=0.05$, and the RWD $=0.13$.  The right panel shows the time averaged line profile (black) and the RMS profile (blue).  The animated version of this figure also plots the line profile changing with $\tau$ (gray), which is in-sync with the bottom panel and isodelay surfaces on the left. The profiles are each normalized to their respective maximum amplitude and binned to 100\,km\,s$^{-1}$ per bin.}  
\label{fig:sphere:s-1p0}
\end{figure} 

In Figure~\ref{fig:sphere:s-1p0} we show the 2DRF for an ICE BLR model that has a spherical geometry and a radiation pressure driven outflow, with gas density and cloud distributions $n(r)\propto r^{-1}$ and $N(r)=constant$ (defined by $s=-1$ and $p=0$), respectively. The left panel shows the parabolic isodelay surfaces corresponding to the continuum pulse at various delays (Equation~\ref{eq:isodelay}) as it propagates through the BLR.  The right panel shows the 2DRF, 1DRF, average line profile, and RMS line profile.  At $\tau=0$, only clouds located along the LOS from the center of the BLR to the observer respond and these have the largest blueshifted velocities. The strongest changes in the line profile are shown by the RMS profile, which is peaked and skewed to the blueshifted side of the line (far-right sub-panel). The response remains entirely blueshifted until $\tau \geq 1/2Y_{BLR} = 0.025$ (Equation~\ref{eq:isodelay} with $r=R_{in},\, \Theta = 90^o$), when the response-front crosses into the far-side hemisphere.  The response covers the greatest spread in LOS velocity at the light-crossing time of the inner radius, $\tau=1/Y_{BLR}=0.05$ (bold curve in left panel), when it reaches the maximally redshifted LOS velocity. This also corresponds to the width of the flat-topped portion of the 1DRF (bottom-right sub-panel). The response then narrows in velocity space as $\tau$ increases and becomes entirely redshifted for $\tau\geq 0.5$; the instantaneous line profiles are now red-asymmetric.  After crossing the inner cavity, the response amplitude decays as $\tau$ increases until the BLR light-crossing time is reached ($\tau=1.0$). It can be seen from Equation~\ref{eq:ana:tf} that the relatively rapid decay is a consequence of the steeply declining emissivity distribution, where $\varepsilon_V \propto r^{-10/3}$ (Equation~\ref{eq:emiss}), in this case. Although changes in the line profile begin in the blue wing then propagate redward as $\tau$ increases, the time-averaged line profile is symmetric, since in this ICE model there is an equal amount of line emission from the redshifted and blueshifted hemispheres of the BLR.  

\begin{figure}
    \centering
    \includegraphics[trim=0 160 0 150, clip, width=\columnwidth]{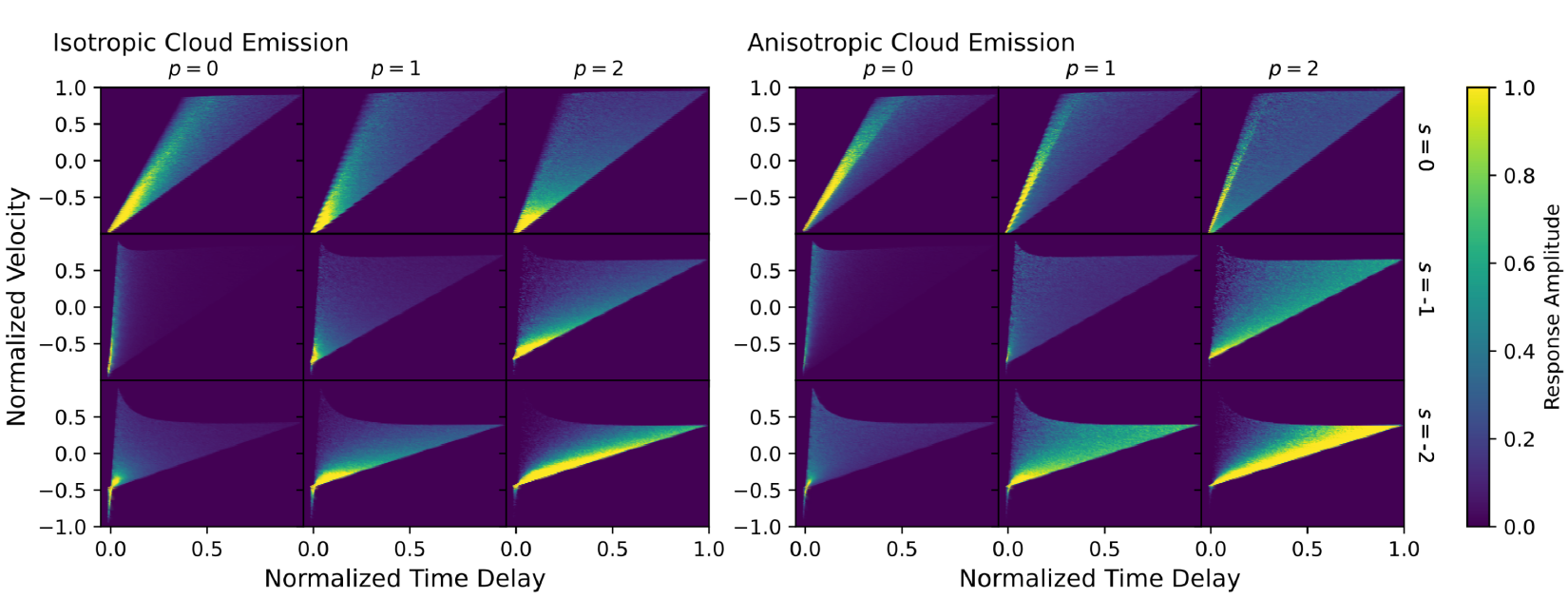}
    \caption{Velocity-resolved response functions (2DRFs) of spherical BLRs with radial outflow for varying $s$ and $p$ model parameters.  The velocities are normalized to the maximum velocity of the respective BLR models. For instance, the response for $s=0$, $p=0$ is normalized to $17\times 10^3$\,km\,s$^{-1}$ and for the $s=-2$, $p=2$ model, to $7\times 10^3$\,km\,s$^{-1}$.  Refer to Figure~\ref{fig:vel} for the maximum velocities in other cases.}
    \label{fig:sphere:2DTFs}
\end{figure}

\begin{figure}[ht]
    \centering
    \includegraphics[trim=0 160 0 150, clip, width=\columnwidth]{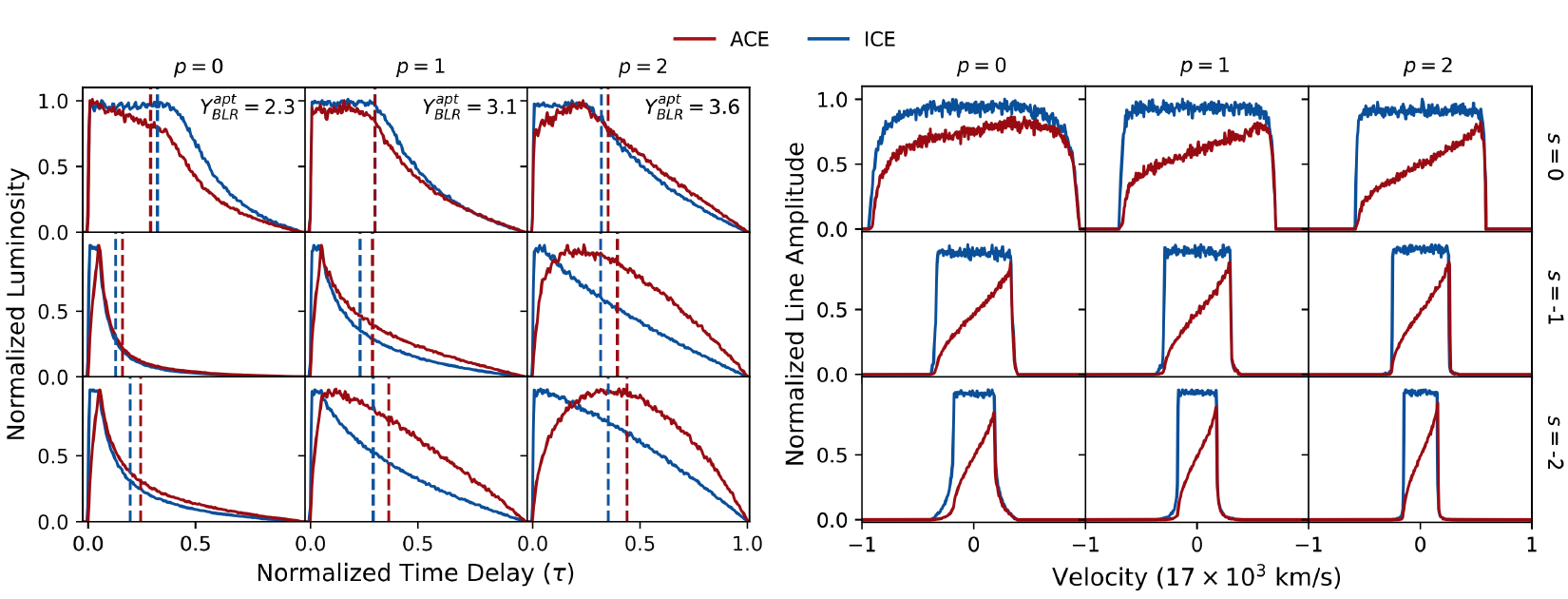}
    \caption{Left: The velocity integrated, 1DRFs for the models shown in Figure~\ref{fig:sphere:2DTFs} for ICE (blue) and ACE (red) models.  The dashed blue and red lines indicate the RWDs for the ICE and ACE models, respectively.  The values of $Y_{BLR}^{apt}$ are given for each $p$ when the ensemble includes matter-bounded clouds (see Section~\ref{sec:res:sphere:mbc} for details).  Right: The delay time integrated, line profiles of Figure~\ref{fig:sphere:2DTFs}.  The profiles of ACE models (red) are normalized to the peak of the ICE model profiles.  For each model, the velocities are normalized to the maximum velocity of the broadest line, which is when $s=0$ and $p=0$.}
    \label{fig:sphere:TFLP}
\end{figure}

The 2DRFs of spherical shell BLR models, with $p=0,~1,~2$ and $s=0,~-1,~-2$ are shown in Figure~\ref{fig:sphere:2DTFs} for both ICE and ACE cases. Their corresponding 1DRFs and delay-averaged line profiles are shown in the left and right panels of Figure~\ref{fig:sphere:TFLP}, respectively.  Considering first the ICE models with $s=-1$ and $-2$ (middle and bottom rows of Figures~\ref{fig:sphere:2DTFs} and Figure~\ref{fig:sphere:TFLP}), it can be seen that as $p$ is increased from $p=0$ to $p=2$ the response narrows in velocity space and it's amplitude declines more gradually with increasing $\tau$. For the larger values of $p$, the clouds with the highest velocities are located near $R_{in}$ (see Figure~\ref{fig:vel}) and make-up a small fraction of the ensemble. These clouds, therefore, make a relatively small contribution to the response, which is dominated by clouds at larger radii, resulting in a response that is quite narrow in velocity-space. The more gradual decline in the tail ($\tau>1/Y_{BLR}$) of the 1DRF is due to the more gradual decrease in emissivity with $r$, where for $s=-2$, $p=0,~1,~2$, $\varepsilon_V \propto r^{-8/3},~r^{-5/3}$, and $r^{-2/3}$.    

\subsubsection{Matter-Bounded Clouds}
\label{sec:res:sphere:mbc}
\begin{table}
\caption{Statistics for matter-bounded clouds in the $s=0$ models. The first columns is the cloud distribution index parameter $p$. The $2^{nd}-4^{th}$ columns are, respectively, the percentage of matter-bounded clouds, the distance at which the clouds are radiation-bounded, and scaled radial depth, $R_d/R_{MBC}$, prior to continuum pulse initiates.  The $5^{th}-7^{th}$ columns are, respectively, the percentage of radiation-bounded clouds that become matter-bounded, the greatest distance the bound state transition occurred, and the apparent scaled radial BLR size after the response completes.  The true inner radius is $6.17\times10^{16}$\,cm and scaled size $Y_{BLR}=20$, for all BLR models presented here.}
\begin{tabular}{c|c|c|c|c|c|c}
    $p$ & $\%$ of M.B.C. & $R_{MBC}/R_{in}$ & $Y^{mbc}_{BLR}$ & $\Delta$ M.B.C. & $R^{apt}_{MBC}/R_{in}$ & $Y^{apt}_{BLR}$\\
    \hline
    \hline
    0 & $34\%$ & 7.13 & 2.60 & $28\%$ & 10.5 & 2.27\\
    1 & $7.51\%$ & 5.51 & 3.41 & $13\%$ & 7.78 & 3.07\\
    2 & $1.35\%$ & 3.89 & 4.23 & $4\%$ & 6.48 & 3.65\\
\end{tabular}
\label{tab:mb:sphere}
\end{table}

The 1DRF's of the ICE models in Figure~\ref{fig:sphere:TFLP} have a flat-topped portion at short delays, but the widths vary greatly between the $s=0$ and $s=-1$ models. Given that $n(R_d)=10^9$\,cm$^{-3}$ and that cloud sizes are constant when $s=0$, a portion of the clouds are matter-bounded in the BLR's initial state, prior to the onset of the continuum pulse, and thus do not respond to the increase in the ionizing flux (see Equation~\ref{eq:lum}).  There is a transition from matter- to radiation-bounded clouds at the distance $R_{MBC}$, when $d_s=d_c$. Since $R_{MBC} > R_{in}$, the BLR responds as if it has a radial depth $Y_{BLR}^{mbc}=R_d/R_{MBC}$, which is smaller than the true value, $Y_{BLR}$.  The transition radius varies with $p$ and is listed in Table~\ref{tab:mb:sphere} along with the fraction of matter-bounded clouds during the initial state. Even a small matter-bounded cloud fraction corresponds to an $R_{MBC}$ that is several times larger than the true $R_{in}$. For example, when $p=2$, only $\sim 1\%$ of clouds are matter-bounded and don't respond, but $Y_{BLR}^{mbc}\approx4$. 

\begin{figure}[ht]
   \centering
    \includegraphics[width=0.75\columnwidth]{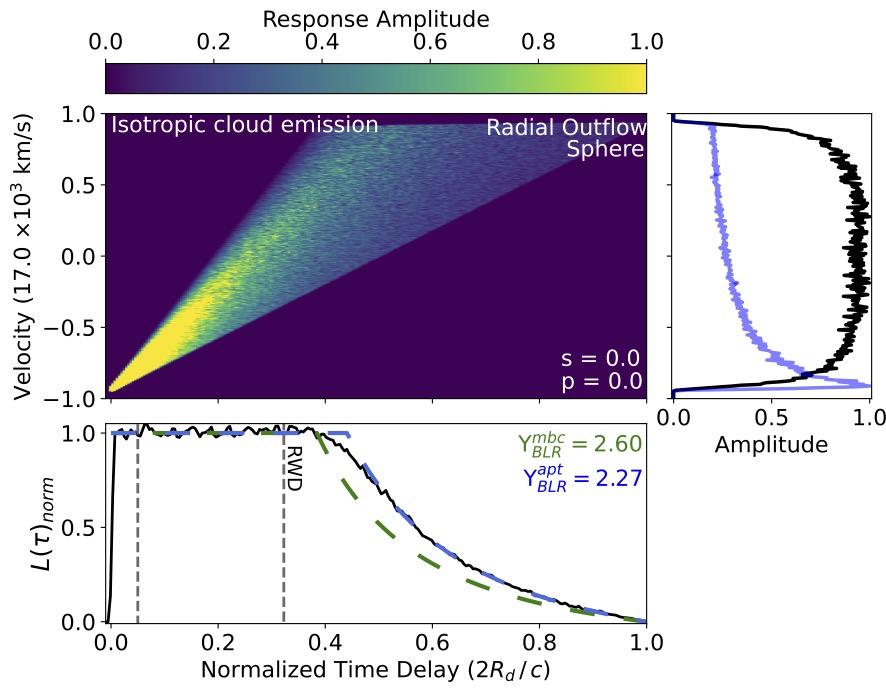}
    \caption{Same as right panel in Figure~\ref{fig:sphere:s-1p0}, but with a gas density index $s=0$ and therefore the ensemble includes matter-bounded clouds.  The green and blue lines are the analytical 1DRFs for $Y_{BLR}^{mbc}=2.60$ and $Y_{BLR}^{apt}=2.27$, respectively.  
    }
    \label{fig:sphere:s0p0}
\end{figure}

In addition to the clouds initially emitting at full capacity, there are clouds with $d_s\lesssim d_c$ that are almost matter-bounded and become matter-bounded when the response-front reaches them. These clouds are at distances $r>R_{MBC}$ and cause a turn-over feature in the response. Over the course of the response, the fraction of radiation-bounded clouds that transition to matter-bounded for $p=0,~1$, and 2 are $28\%,~13\%,$ and $4\%$, respectively.  The clouds transitioning to the matter-bounded state further increases the apparent inner BLR radius to $R_{MBC}^{apt}$, which increases the width of the flat-topped portion of the 1DRF to $1/Y_{BLR}^{apt}$.  For example, Figure~\ref{fig:sphere:s0p0} shows the 2DRF, 1DRF, and averaged line profile for $p=0$ and $s=0$.  Overplotted on the 1DRF are two analytical 1DRFs calculated using Equations~\ref{eq:ana:tf:sphere1} and~\ref{eq:ana:tf:sphere2} with $Y_{BLR}^{mbc}=2.60$, determined from the transition radius $R_{MBC}$, and $Y_{BLR}^{apt}=2.27$, which was determined by $R^{apt}_{MBC}$.  Although these values are similar, the analytical curve for $Y_{BLR}^{apt}=2.27$ is clearly a better match to the 1DRF.  The values of $Y_{BLR}^{mbc}$ and $Y_{BLR}^{apt}$ for the other BLR models are listed in Table~\ref{tab:mb:sphere}.   

\subsubsection{Anisotropic Cloud Emission}
\label{sec:res:sphere:ace}

Figures~\ref{fig:sphere:2DTFs} and \ref{fig:sphere:TFLP} also show the 2DRFs, 1DRFs and $\tau$-averaged line profiles for the ACE models, where $\varepsilon_V$ also depends on $elf$ and therefore becomes a function of $\theta$ and $\phi$, as well as $r$.  For ACE models in which all the clouds are radiation-bounded (as is the case for $s=-1$ and $-2$), the observer sees $<50\%$ of the illuminated face for the clouds on the near-side of the BLR.  Therefore, at short time delays, the blue-wings of the line profiles are suppressed and the 1DRF is no longer flat-topped. The response peak occurs at delays when the \textit{majority} of clouds on the far-side are along the LOS (i.e., $elf>1$) are responding, that is, when $\tau =1/Y_{BLR}$ for $p=0$ and at later delays for the $p=1$ and 2 models, where the clouds are more uniformly distributed.  The $\tau$-averaged line profiles of the ACE models are highly asymmetric and have half the amplitude of the corresponding ICE profiles.  Clouds that are matter-bounded are assumed to emit isotropically and, therefore, have their $elf$ values set to 1. The models that include matter-bounded clouds (which occurs for $s=0$) therefore have 1DRFs that are more similar to their ICE counterparts.  In the line profiles, the asymmetry between the red and blue wings is smallest in the $p=0$ model, which is the model having the highest fraction of matter-bounded clouds, $\sim 34\%$.  The largest asymmetry occurs in the $s=0,~p=2$ model, which has the lowest fraction of matter-bounded clouds of $\sim1\%$.

\subsubsection{The Response Weighted Delay and Luminosity Weighted Radius}
\label{sec:res:sphere:rwd}

\begin{figure}[ht]
\centering
\includegraphics[trim = 0 100 0 100, clip, width=\textwidth]{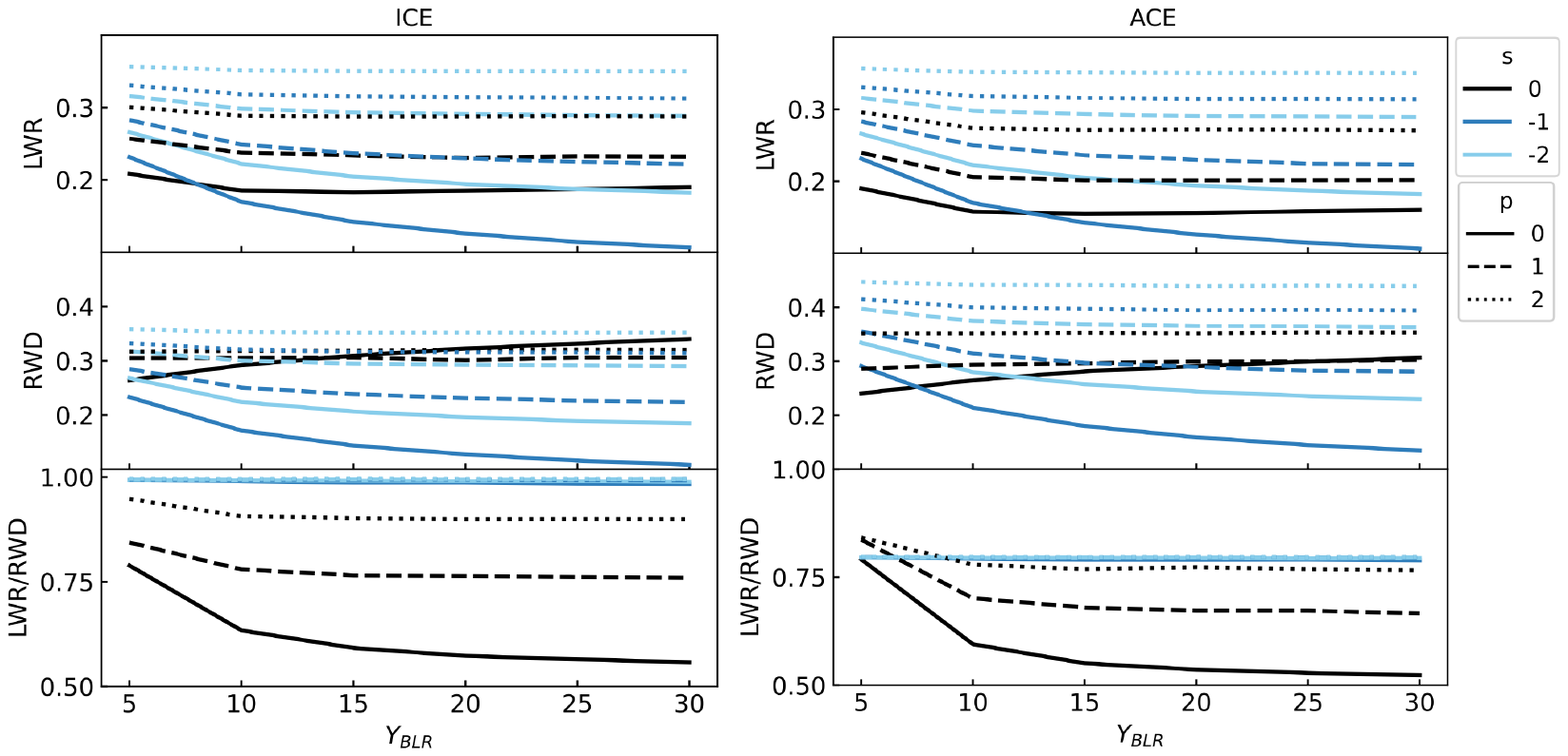}
\caption{The LWR and RWD of the spherical ICE (left) and ACE (right) BLR models of varying $Y_{BLR}$ for each combination of $s$ and $p$ parameters, with their ratio in the bottom panel.}
\label{fig:sphere:LWR}
\end{figure}

Figure~\ref{fig:sphere:LWR} shows the variation with $Y_{BLR}$ of the LWR, RWD, and the ratio LWR/RWD for ICE and ACE BLRs, and for each combination of $s$ and $p$.  

For ICE models with only radiation-bounded clouds ($s\neq0$), the LWR and RWD both decrease with increasing $Y_{BLR}$, similar to the analytical relations shown in Figure~\ref{fig:ana:LWR}, and the ratio LWR/RWD $=1$.  Additionally, since LWR is not dependent on $elf$, the LWR values for the ICE and ACE models are identical.  However, the response peaks at longer delays for ACE models, as discussed in Section~\ref{sec:res:sphere:ace}, and thus the RWD is greater than for the corresponding ICE models.  Therefore, for ACE models with $s\neq0$, LWR/RWD $\approx0.8$ for all $p$ and $Y_{BLR}$ values.   

For models with $s=0$, the RWD \emph{increases} with $Y_{BLR}$ due to the presence of non-responding, matter-bounded clouds. The largest increase in RWD occurs for $s=0$, $p=0$, when about $\sim 30\%$ of the cloud ensemble is matter-bounded. As a result, LWR/RWD decreases as $Y_{BLR}$ increases, implying that the RWD overestimates the LWR and is dependent on the fraction of matter-bounded clouds and hence also $Y^{apt}_{BLR}$).  The overestimation of LWR due to the presence of matter-bounded clouds ranges by factors of $\sim1.1$ to nearly $2$ for the ICE models.  Furthermore, the LWRs for respective ICE and ACE models are not equivalent because the matter-bounded clouds have $elf=1$ (emit isotropically).  Therefore, when $s=0$, the LWRs for ACE models are slightly lower than the LWRs of corresponding ICE models because the clouds emitting isotropically in the ensemble are located within $R^{apt}_{MBC}$.  The combined effect of ACE and matter-bounded clouds means the RWD overestimates the LWR by factors of $\sim1.2$ to $\sim2$.  

\subsection{Keplerian Disk BLR}\label{sec:res:disk}

Here we present the results of models in which the BLR is configured as a Keplerian disk. As specific examples, we discuss the reverberation responses for the ICE and ACE cases for an edge-on ($i=90^o$), rotating thin disk ($\sigma=15^o$), where the gas density decreases as $r^{-2}$, $U$ is constant ($s=-2$), cloud size increases with distance, and the cloud radial distribution is constant ($p=0,~N(r)$ constant).  In this case, all of the clouds are radiation-bounded.  

\begin{figure}[t]
%\begin{interactive}{animation}{Figures/tormac_BLR_CF0.3_off__sharp__p0_90_20_15_100.0k_sneg2_KDiskn8_ice_tri.mp4}
\includegraphics[width=\textwidth]{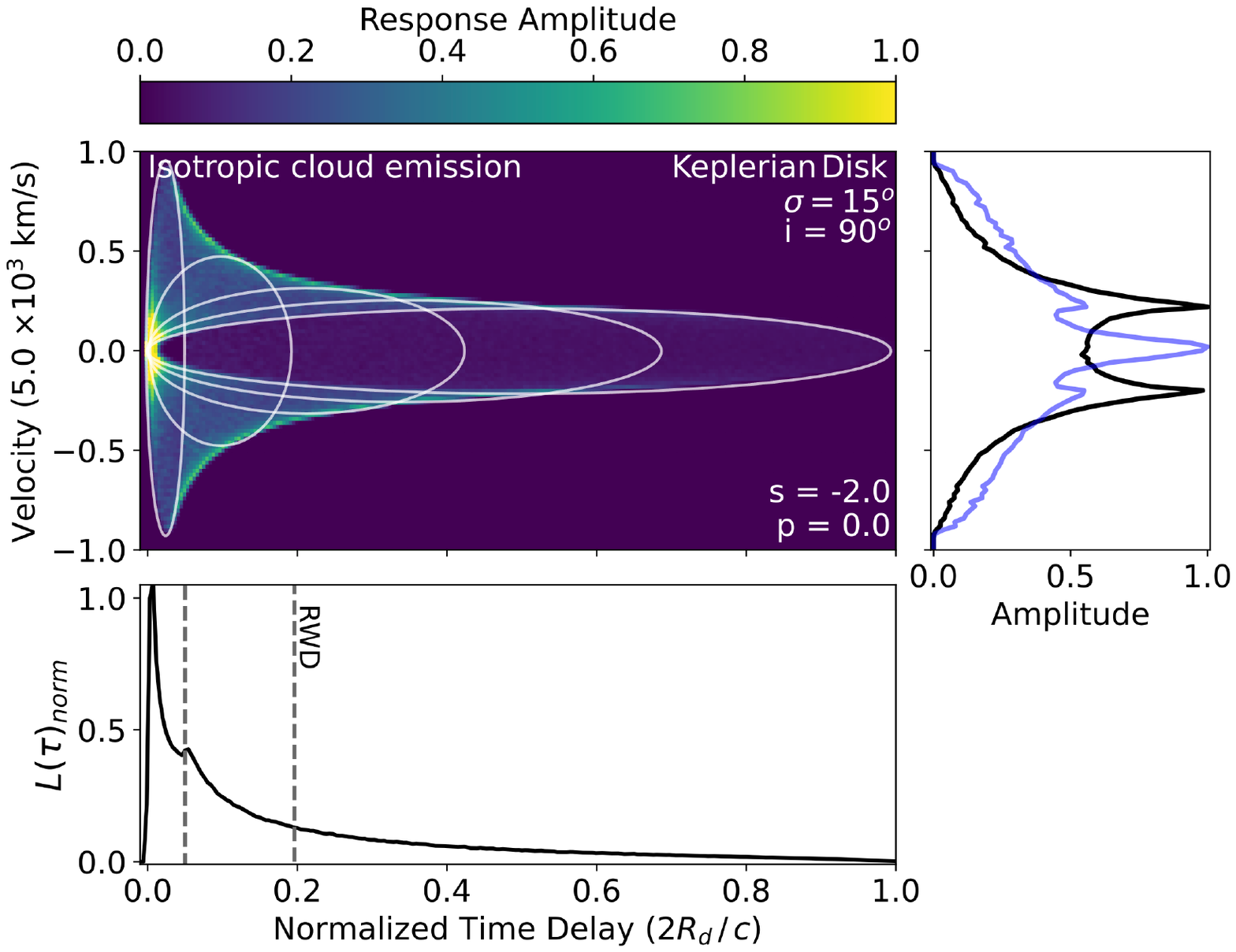}
%\end{interactive}
    \caption{Same as the right panel in Figure~\ref{fig:sphere:s-1p0}, but for a disk-like ($\sigma=15^o$) BLR with clouds moving in Keplerian orbits.  The disk is viewed edge-on ($i=90^o$), $s=-2$, and $p=0$.  The dashed lines indicate the light-crossing time of $R_{in}$ and the measured RWD $=0.197$.  The white ellipses are defined by Equation~\ref{eq:ellipse}. The system is entirely radiation-bounded clouds.  This figure has an animated version that is available in the online version of this paper, which does not display the ellipses.  In the animation, the 1DRF is plotted with $\tau$ from 0 to 1, along with the $\tau-$varying line profile (gray). The changing line profile is a narrow, bright peak at $\tau=0$, then quickly becomes lower in amplitude and broader with double peaks at the extremities of the profile's wings.  These low-amplitude peaks slowly move back towards the center of the profile until $\tau=1$. The $\tau-$averaged line profile (black), RMS (blue), and, in the animation, the $\tau-$varying line profiles (gray) are normalized to their respective maxima.}
    \label{fig:disk:ice}
\end{figure}

%ICE
Figure~\ref{fig:disk:ice} displays the 2DRF, 1DRF, time-averaged, and RMS line profiles of the ICE model. For a specific orbital radius $r$, the projection of the isodelay surfaces and orbital velocities in the disk's plane is an ellipse in the 2DRF \citep[Equation~\ref{eq:ellipse};][]{Perez1992TheFunctions}.  As can be seen in Figure~\ref{fig:disk:ice}, the characteristic rocket or bell-shaped 2DRF is created by the superposition of ellipses of varying $r$.  Clouds at $R_{in}$ have the largest LOS velocity range and respond at a shorter time delay than the clouds located at  larger radii.  The major axis of the corresponding ellipse is therefore aligned with the velocity axis. Conversely, the ellipse associated with the clouds at $R_d$, has the smallest velocity range and spans the largest range in time delay, and hence it's major axis is aligned with the time delay axis.  Rather than the flat-top that was seen for the 1DRF of the spherical BLR, the disk 1DRF has a sharp peak at $\tau = 0.0$, then a second, weaker peak in the response at the light-crossing time of the inner radius, $\tau=0.05$, due to the decrease of clouds while the response-front crosses the inner cavity.  In this model, $\varepsilon_V\propto r^{-8/3}$ and as $\tau$ increases, more clouds at larger radii respond and, the tail of the 1DRF smoothly decreases until the light-crossing time of the BLR is reached ($\tau = 1$).  For an edge-on rotating disk, the $\tau-$averaged line profile has distinctive, symmetrical redshifted and blueshifted peaks since the emission from the receding and approaching sides of the disk contributes equally.  The response begins at $v_{rot}^{||}\approx0$\,km\,s$^{-1}$ and the amplitude is large enough that the RMS profile has a central peak.  The response then moves quickly and simultaneously to each wing,  contributing to the weaker, symmetrical peaks in the RMS profile.    

\begin{figure}
%\begin{interactive}{animation}{Figures/tormac_BLR_CF0.3_off__sharp__p0_90_20_15_100.0k_sneg2_KDiskn8_tri.mp4}
\includegraphics[width=\textwidth]{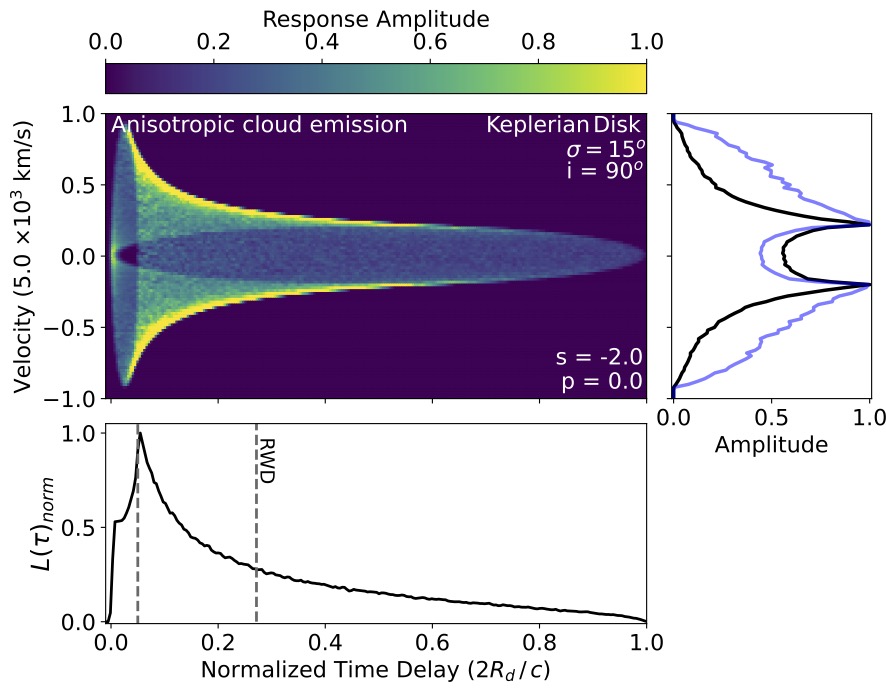}
%\end{interactive}
\caption{The same as Figure~\ref{fig:disk:ice}, including an animated version that is available online, but for an ACE model.  In the animation, the $\tau-$varying line profile is a narrow, low-amplitude peak at $\tau=0$, then quickly becomes brighter in amplitude and broader with double peaks at the extremities of the profile's wings.  These high-amplitude peaks slowly move back towards the center of the profile until $\tau=1$.
\label{fig:disk:ace}}
\end{figure}

The effect of ACE decreases the response amplitude during short time delays because the observer sees very little of the illuminated faces of the responding clouds.  The initial response amplitude is diminishes by almost half compared to the amplitude in the ICE 1DRF model in Figure~\ref{fig:disk:ice}.  Instead, in the ACE model, the response peaks at $\tau=0.05$, when the observer receives the response from the clouds with $elf>0.5$.  For the same reason, although the response begins in the core of the line profile, the effect of ACE decreases the response amplitude and the RMS profile doesn't have a central peak. 

\begin{figure}[t]
    \centering
    \includegraphics[trim=0 160 0 150, clip, width=\columnwidth]{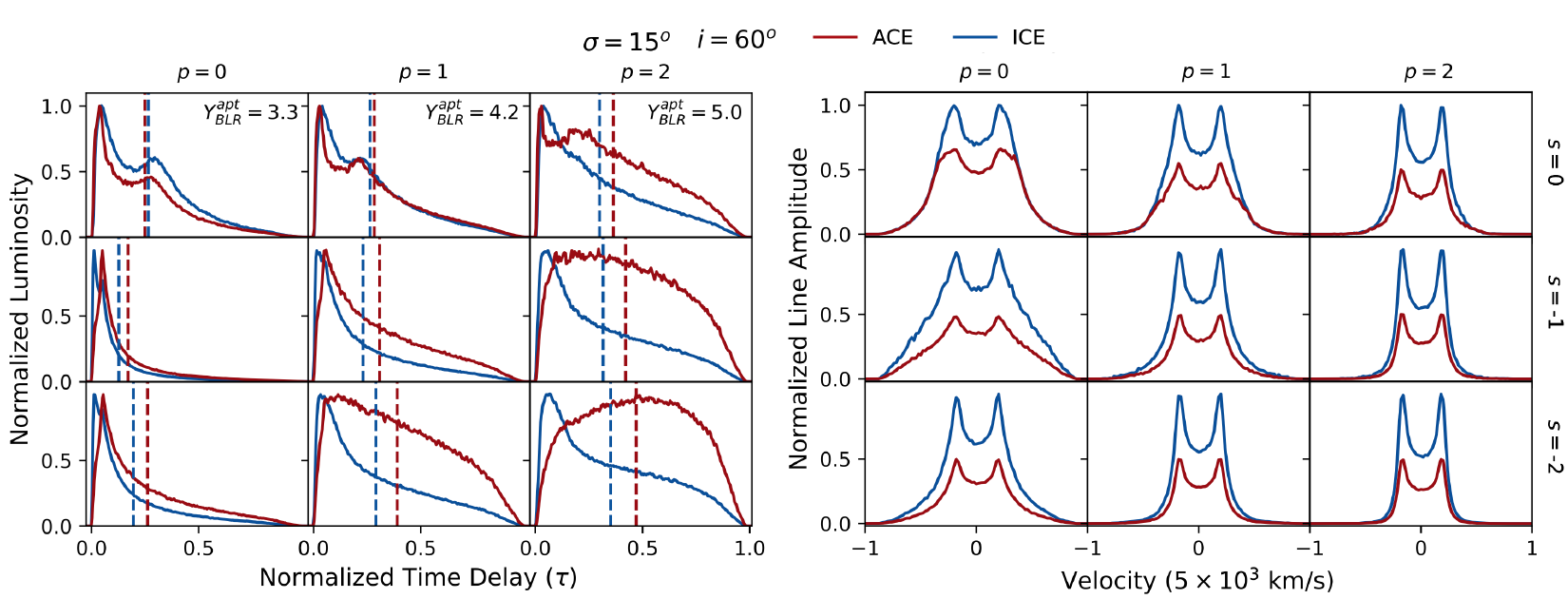}
    \caption{The same as Figure~\ref{fig:sphere:TFLP}, but for disk-like ($\sigma=15^o$) BLRs with clouds moving in Keplerian orbits and viewed from $i=60^o$.  
    }
    \label{fig:disk:TFLP}
\end{figure}

Figure~\ref{fig:disk:TFLP} shows the 1DRFs and line profiles for disk models inclined at $i=60^o$ with varying values of the $s$ and $p$ parameters for both ICE and ACE cases.  We chose to present the $i=60^o$ BLR models, since an edge-on BLR would be obscured to an observer by the surrounding dusty torus.  However, we still want to present a high inclined disk to show distinctly separated peaks in the line profiles.  Other choices of intermediate inclinations produce qualitatively similar results.  When $s=-1$ and $-2$, all of the clouds are radiation-bounded.  For these cases of ICE models, $\varepsilon_V(r)$ decreases less rapidly as $s$ decreases and $p$ increases, therefore the response decays more gradually with $\tau$.  In the ACE models, the clouds on the far-side relative to an observer contribute more emission to the total response, so the peak response occurs at longer time delays as $p$ increases.  When clouds are concentrated near $R_{in}$ (i.e., $p=0$), the peak is at $\tau=1/Y_{BLR}$. 

Both the ICE and ACE models have symmetrical line profiles with redshifted and blueshifted peaks, characteristic of rotating disks.  Since the $elf$ is symmetrical about the disk axis, the ICE and ACE line profiles have the same shape, but the ACE model profiles have half the amplitude. 

\begin{table}[b]
\caption{The same as Table~\ref{tab:mb:sphere}, but for a disk geometry with $i=60^o$.}
\begin{tabular}{c|c|c|c|c|c|c}
    $p$ & $\%$ of M.B.C. & $R_{MBC}/R_{in}$ & $Y^{mbc}_{BLR}$ & $\Delta$ M.B.C. & $R^{apt}_{MBC}/R_{in}$ & $Y^{apt}_{BLR}$\\
    \hline
    \hline
    0 & $22.87\%$ & 5.35 & 3.73 & $20\%$ & 7.46 & 3.3\\
    1 & $3.72\%$ & 3.89 & 5.12 & $6.6\%$ & 5.67 & 4.2\\
    2 & $0.43\%$ & 3.24 & 6.15 & $1.4\%$ & 4.70 & 5.0 \\
\end{tabular}
\label{tab:mb:disk}
\end{table}

When $s=0$, a portion of the clouds are matter-bounded prior to the onset of the continuum pulse and hence do not respond.  Consequently, both peaks in the 1DRF are broader and the second occurs at $\tau\approx1/Y_{BLR}^{apt}$.  The percentages of matter-bounded clouds, $R_{MBC}$, and $Y_{BLR}^{apt}$ for each model are listed in Table~\ref{tab:mb:disk}.  Since $elf=1$ (isotropic emission) when a cloud is matter-bounded, the ratio of the ICE and ACE line profiles amplitudes approaches 1 as the fraction of matter-bounded clouds increases (see Table~\ref{tab:mb:disk}).

\begin{figure}[ht]
    \centering
    \includegraphics[trim=0 110 0 110, clip, width=\columnwidth]{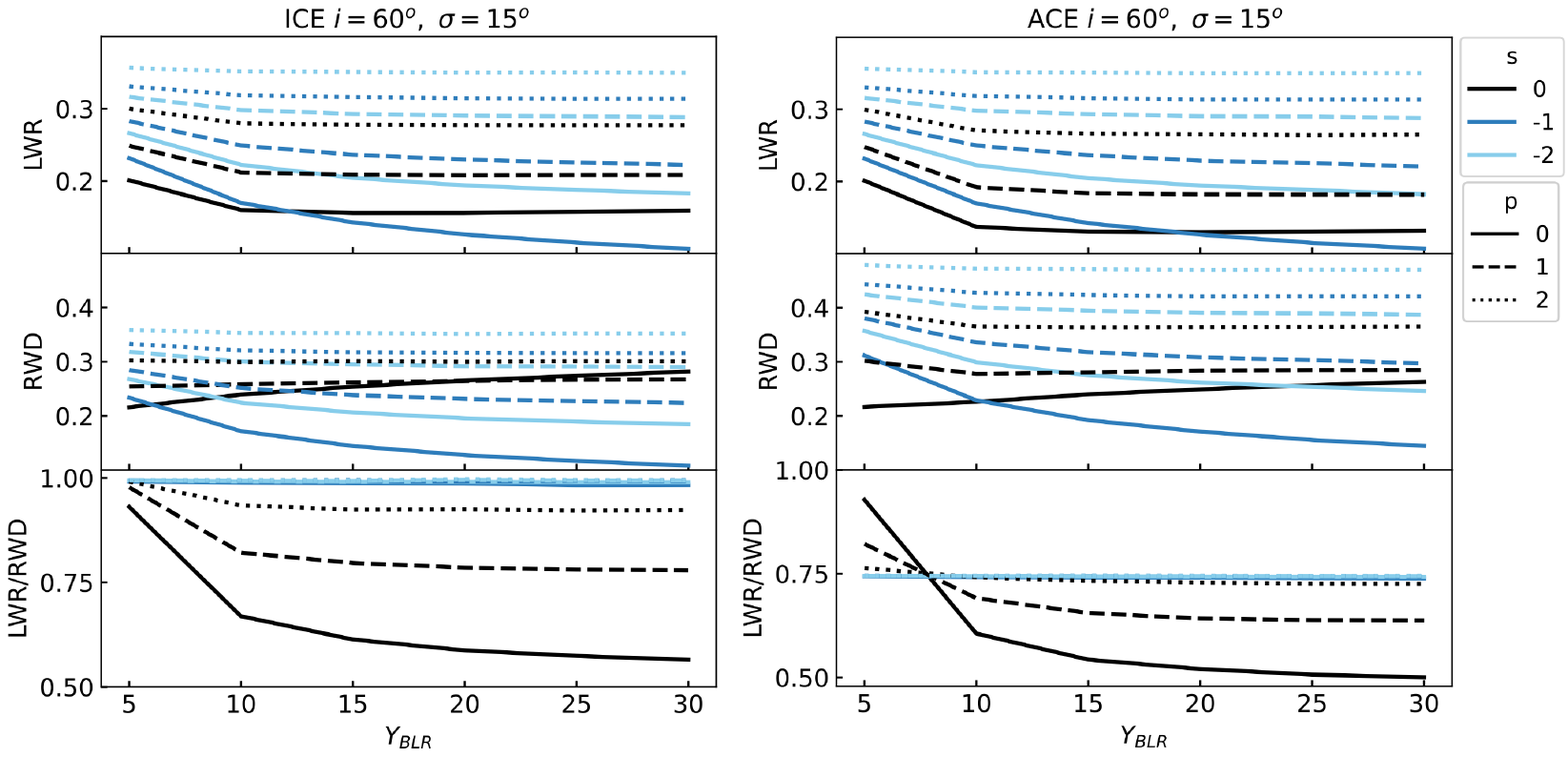}
    \caption{The same Figure~\ref{fig:sphere:LWR}, but a rotating disk with $i=60^o$.}
    \label{fig:disk:LWR}
\end{figure}

% LWR, RWD
The LWRs and RWDs for a disk with $i=60^o$, shown in Figure~\ref{fig:disk:LWR}, show the same general trends with $Y_{BLR}$ as the spherical BLR.  In particular, the ICE models (left panel) and the corresponding spherical BLR models (left panel in Figure~\ref{fig:sphere:LWR}) show very similar behavior.  However, the RWDs for the ACE models are longer for a disk than a sphere and therefore the LWR/RWD ratio has slightly lower values (LWR/RWD $=0.75$, for $s\neq0$).  When $s=0$, LWR/RWD $<1$ for both ICE and ACE models, and decreases as the fraction of matter-bounded clouds increases ($p=2\rightarrow0$) and as $Y_{BLR}$ increases.  The ACE models have consistently lower LWR/RWD values than the corresponding ICE.  More details on the results of models with matter-bounded clouds is in Section~\ref{sec:res:disk:mbc}. 

\subsubsection{Effects of Inclination}

\begin{figure}[t]
    \centering
    \includegraphics[width=\columnwidth]{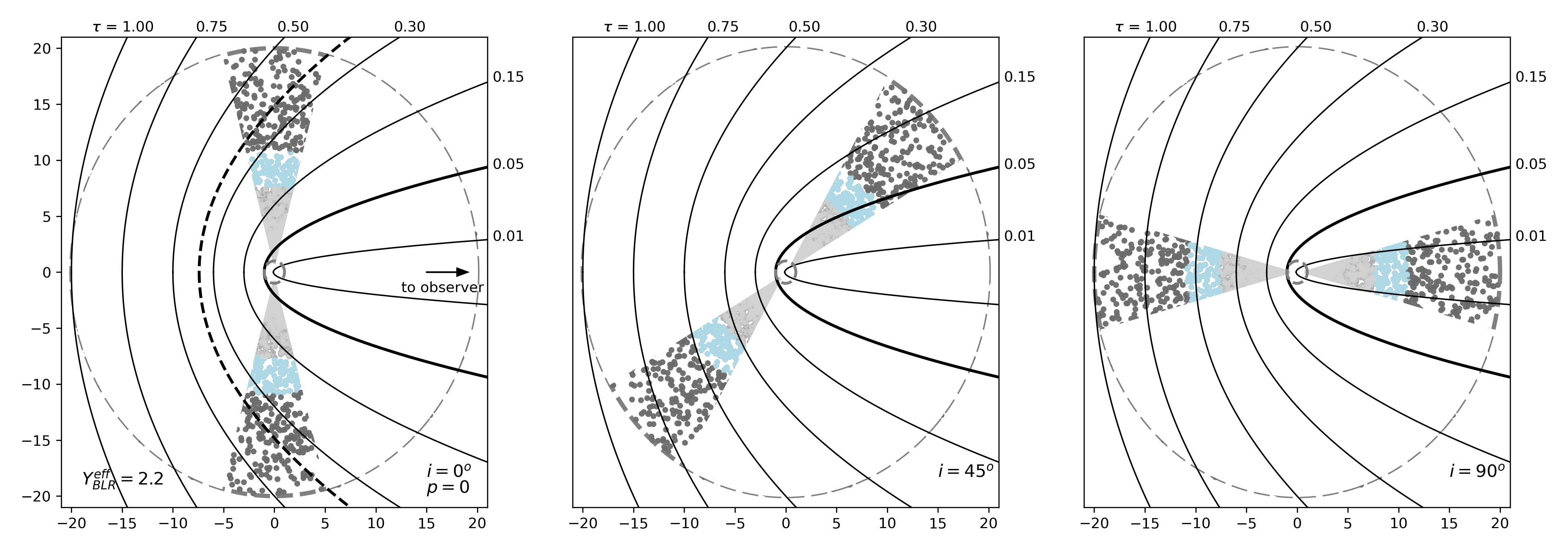}
    \caption{Isodelay surfaces over plotted for a disk-like, $\sigma=15^o$, BLR viewed from left to right at, $i=0^o$, $i=45^o$, and $i=90^o$.  The isodelay surfaces are labelled with the corresponding delays, $\tau$, across the top and down the right-hand side of each sub-figure.  When $s=0$, the light gray points are matter-bounded clouds, that don't respond.  The light blue points represent clouds become matter-bounded when the continuum pulse reaches them and the gray clouds are always radiation-bounded.  If $s\neq0$, then all of the clouds in the ensemble are radiation-bounded.  The dashed line in the $i=0^o$ panel is the isodelay surface corresponding to inflection point seen in the response in the left panel of Figure~\ref{fig:disk:i}.  The thicker, black line is the light-crossing time for $R_{in}$.}
    \label{fig:iso:disk}
\end{figure}

\begin{figure}
    \centering
    \includegraphics[trim=0 110 0 110, clip, width=\columnwidth]{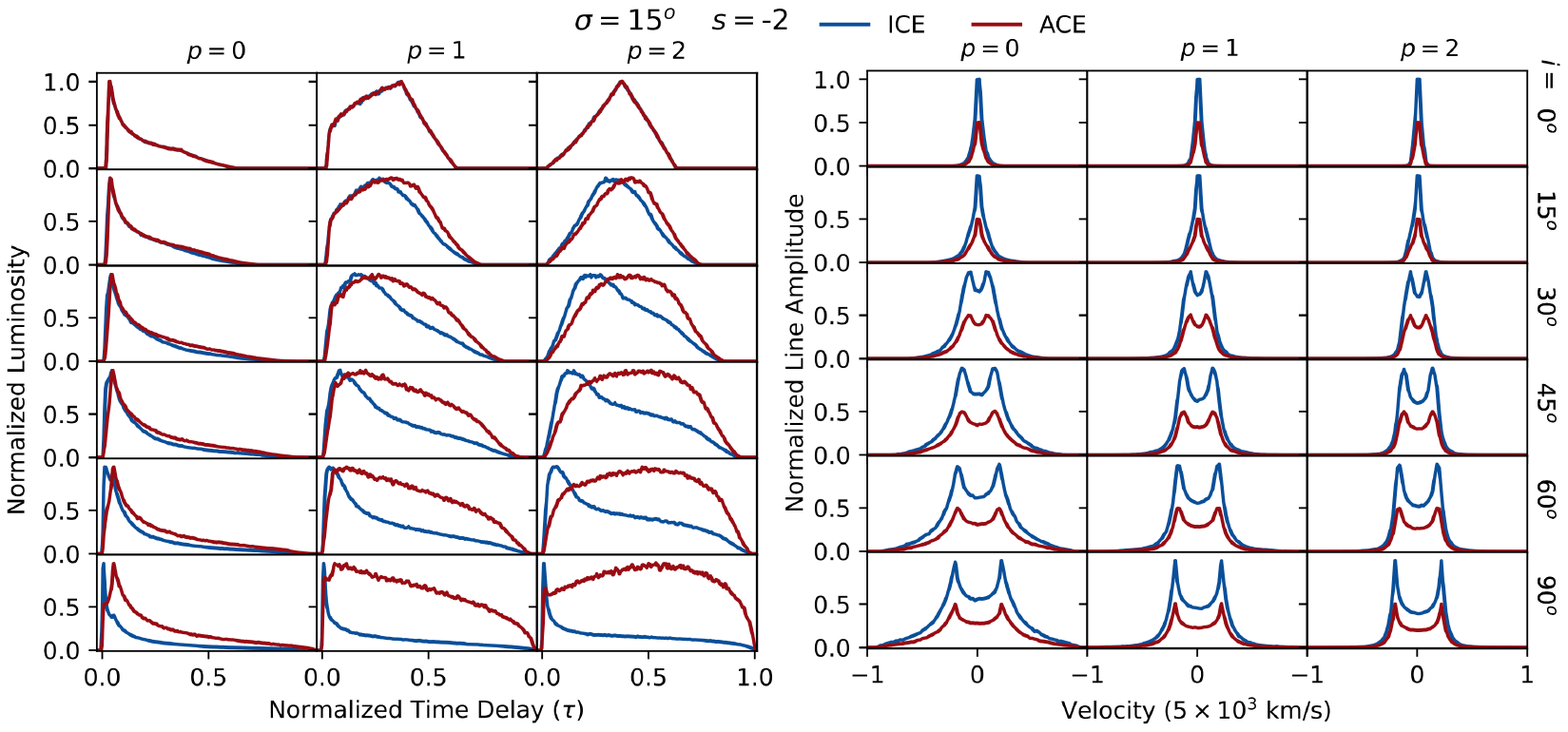}
    \caption{1DTFs (left) and line profiles (right) for a disk-like ($\sigma=15^o$) BLR with clouds moving in Keplerian orbits and viewed from face on to edge-on, $i=0^o,~45^o,~90^o$, for each row and increasing $p=0,~1,~2$ for columns left to right, and $s=-2$.}
    \label{fig:disk:i}
\end{figure}

Cross-sectional diagrams of the disk-like BLR ensembles at $i=0^o,~45^o$ and $90^o$ with isodelay surfaces over plotted are shown in Figure~\ref{fig:iso:disk}.  The effects of inclination on the 1DRFs and line profiles are shown in Figure~\ref{fig:disk:i} for both ICE and ACE models with $s=-2$ and varying $p$.  When the disk is face-on to the observer, the 1DRF is zero until $\tau=0.03$ when the response-front reaches the clouds at the inner-radius.  As $i$ increases, the response begins at progressively shorter delays until, at $i=90^o-\sigma$ the response begins at $\tau=0$.  Additionally for ICE models, the 1DRF peak becomes narrower and the tail overall extends to longer delays, as $i$ increases.  The ACE 1DRFs show much stronger, extended responses at larger inclinations than their ICE counterparts. 

When the disk is face-on, there is a sudden decrease in the 1DRF at $\tau=0.37$. The inflection point corresponds to the delay, $\tau= \frac{1}{2}(1-\cos{\left(90^o-\sigma\right)})$, when the isodelay surface intersects with the outermost edge of the disk surface facing the observer,  shown  in Figure~\ref{fig:iso:disk} as a dashed line.  The 1DRF peaks at this $\tau$-inflection point when $p=1$ and $2$, as the largest number of clouds are responding at this delay.  The response is complete at $\tau= \frac{1}{2}(1-\cos{\left(90^o+\sigma\right)})$. 

The right panel of Figure~\ref{fig:disk:i} shows the line profiles for the same models, where the profile amplitudes have been normalized to the maximum amplitude for the ICE model.  With respect to the line center, the peaks are located at $\pm \sin i\,v_{min}/v_{max}$, therefore the gap between them is zero at $i=0^o$ resulting in a single-peaked profile \citep{Robinson1995OnNuclei}.  Otherwise, when the disk is inclined, the line profile shape has the characteristic, symmetrical peaks about the center of the profile.  The trough between the peaks also gets deeper with increasing $i$ and $p$. 

\subsubsection{Models with Matter-Bounded Clouds}
\label{sec:res:disk:mbc}

\begin{figure}[ht]
    \centering
    \includegraphics[trim=0 110 0 110, clip, width=\columnwidth]{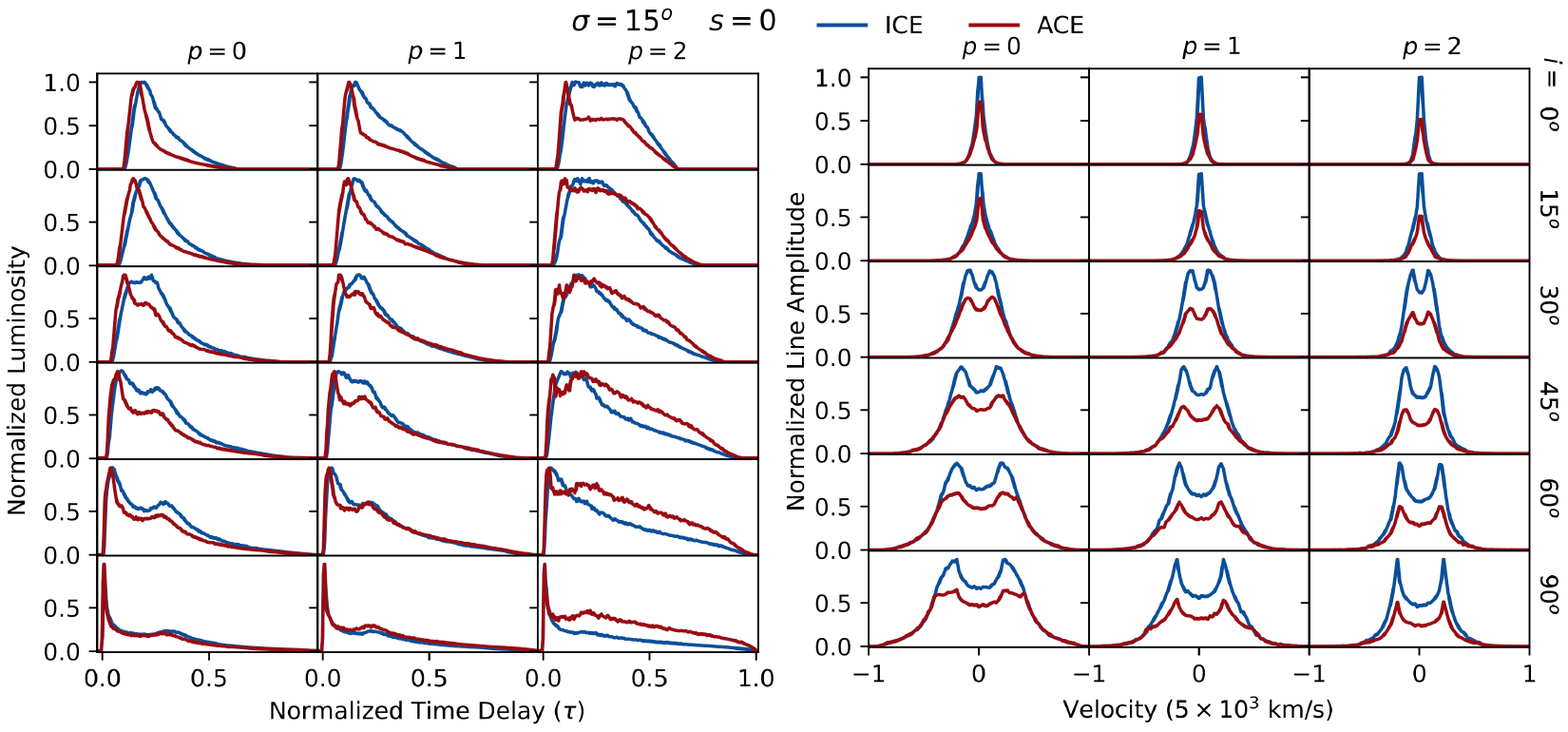}
    \caption{The same as Figure~\ref{fig:disk:i}, but with matter-bounded clouds in the ensemble ($s=0$).}
    \label{fig:disk:i:mbc}
\end{figure}

Figure~\ref{fig:disk:i:mbc} is the same as Figure~\ref{fig:disk:i}, but for models with $s=0$.  Here, clouds within $R_{MBC}$ are matter-bounded and do not respond.  This creates an ``apparent" inner cavity in the BLR that is larger than $R_{in}$.  Therefore, the response has a delayed start that is dependent on $R^{apt}_{MBC}$, in addition to its dependence on $i$ and $\sigma$, as discussed above. However, as $R_{MBC} < R_d$, this does not affect the clouds at the largest radii in the BLR and the response ends at the same delay, $\tau= \frac{1}{2}(1-\cos{\left(90^o+i+\sigma\right)})$, regardless or whether or not matter-bounded clouds are present.  The 1DRFs are double-peaked at $i\geq30^o$, where each peak represents the response-front encountering the near-side then far-side clouds at $R_{MBC}$.  The time-delay between the two peaks lengthens as $i$ increases, since the second peak occurs at $\tau=(2Y^{mbc}_{BLR})^{-1}\left(1-\cos\left(90^o-i\right)\right)$ and also as $p$ decreases due to the increase in the fraction of matter-bounded clouds.

In general, the ICE and ACE line profiles in the right panel of Figure~\ref{fig:disk:i:mbc} have a similar shape.  However, the presence of matter-bounded clouds creates shoulder-like features in the line profiles of ACE BLR models when $i\geq60^o$.  This is most clearly seen for the model with $i=90^o$ and $p=0$, the symmetrical inflections are a result of matter-bounded clouds within $R_{MBC}$ emitting isotropically, then, as the isodelay surface propagates throughout the BLR, radiation-bounded clouds at larger radii respond and emit anisotropically.  The width of these shoulders relates to $R_{MBC}$ and thus increases as $p$ increases.  The width of the shoulders broadens with increasing $p$ and become more smoothed with decreasing $i$, therefore these features we do not clearly see for $p>1$ and $i\leq45^o$.                      

\section{Discussion}\label{sec:discuss}

We have presented a new BLR reverberation mapping forward-modeling code, BELMAC, that incorporates various velocity field models to calculate the velocity-resolved response.  We assume the clouds are bound spheres, which follows similar modeling work such as \citet{Lawther2018QuantifyingDelays,Goad2012TheTorus,Marconi2009ONINDICATORS,Netzer2008IonizedNuclei,Robinson1990TheVariations}.  The version of the code used here assumes hydrogen recombination to determine cloud luminosity, where the clouds can optionally emit isotropically (ICE) or anisotropically (ACE). We presented BLR cloud ensemble models with a radiation pressure driven outflow in a sphere and Keplerian orbits in a disk, respectively.

BELMAC does not provide a full hydrodynamic treatment of the gas flows and does not consider inter-cloud radiative transfer.  However, representing the BLR as a cloud ensemble, offers the ability to quickly and efficiently explore multiple configurations of BLR geometries and velocity fields.  It also allows for compatibility with photoionization model grids, which will be implemented in the forthcoming version.  This representation may impose some limits on BELMAC’s predictions.  Ideally, reverberation mapping modeling codes would include a detailed treatment of hydrodynamics, incorporating radiative transfer and photoionization physics.  However, these codes do not yet exist and would no doubt be computationally expensive.

\subsection{The Radiation Pressure Driven Flow}

The BLR clouds are photoionized and therefore subjected to radiation pressure forces due to scattering and absorption.  Disregarding radiation pressure could lead to significant underestimations or overestimations of $M_\bullet$ as determined by the virial method \citep[see e.g.,][]{Marconi2008TheGalaxies,Netzer2010THENUCLEI}.  Although Keplerian motion was found to dominate the BLR velocity field in the AGN discussed in Section~\ref{sec:intro:RM}, there may also be weak radial flows.  In fact, there is evidence for inflows in NGC 3783 \citep{Amorim2021A3783}.  Furthermore, these AGN have low to moderate Eddington ratios ($\Gamma<0.2$, based on the reported $L_{AGN}$ and $M_\bullet$), but outflows are expected to be much stronger in high and super-Eddington AGN.  The outflows may be associated with a disk wind, with the relative prominence of the disk and wind components scaling with AGN luminosity \citep{Elitzur2014EvolutionNuclei}.  This suggests that the dominant emission line producing structures in the BLR are dependent on the SMBH's accretion rate.    

In our adopted model (Equation~\ref{eq:eom}), there is an outflow when $\Gamma F_M > 1$ where $F_M \propto 1/N_{col}$, therefore the conditions for an outflow in our models depend on the Eddington ratio, $\Gamma$, and column density.  Since $\Gamma$ is dependent on $L_{AGN}$ and $M_\bullet$, it will be an output when BELMAC is applied to model real data, but here we have fixed the AGN values to $\Gamma=0.07$ and $M_\bullet=10^8$\,M$_\odot$.  To achieve an outflow, the clouds must have $N_{col}<5.26\times10^{22}$\,cm$^{-2}$ (see Equation~\ref{eq:eom}).  For models with $s=0$, clouds have clouds a constant cross-sectional area and $N_{col}(r)\sim10^{22}$\,cm$^{-2}$, thus $\Gamma F_M > 1$ throughout the entire BLR.  When cloud cross-sectional areas vary with distance ($s\neq0$), the clouds with $N_{col}(r) > 5.26\times10^{22}$\,cm$^{-2}$ are too massive for radiative forces to overcome gravity.  As $N_{col}(r)$ decreases the $F_M$ increases, because the clouds expand to maintain constant mass.

The above density calculations are also dependent on the initially-defined gas density, $n(R_d)$.  If $n(R_d)$ is large, e.g., $n(R_d)=10^{10}$\,cm$^{-3}$, $\Gamma F_M<1$ throughout the BLR and $v_{rad}(r)\rightarrow\sqrt{GM_\bullet/r}$.  Furthermore, in these models we have assumed that the mass of the clouds is conserved (i.e., $m_{cl}(r)=$ constant). In real BLRs, the relation between the sizes of clouds and their masses may vary, leading to a more complex gas density distribution.  Varying the cloud mass would alter our adopted radial cloud motion model and could produce less-boxy line profiles. The last parameter to consider is $\alpha_f$, the fraction of $L_{AGN}$ absorbed by a cloud.  A more realistic BLR model would let $\alpha_f$ depend on the ionization parameter and $N_{col}$, but here we have opted to keep that quantity constant ($\alpha_f=0.5$; Equation~\ref{eq:acc:total}).  Future versions of BELMAC will include the radial and time dependence, $\alpha_f(r,t^\prime)$.   

The line profiles produced by the spherical shell-radial flow models tend to be boxy unlike the observed broad line profiles in most AGN, which typically resemble Gaussian or Lorentzian profile. This is a feature of the radial velocity field model adopted here. As can be seen in Figure~\ref{fig:vel}, the radial velocity curves rapidly level off to a terminal velocity, within $\sim 0.1 - 0.3\,R_d$. As $v(r)=$ constant in the outer regions, the dispersion in LOS velocity varies little with radius, resulting in flat topped profiles.  Although, varying $\alpha_f(r,t^\prime)$ could reduce the acceleration so that the terminal velocity is reduced at a larger radius.  Here, the velocity range for the large majority of clouds in the ensemble is quite small, resulting in boxy line profiles that don't resemble most observed broad emission lines.  For $s=0$ and $p=0$, the clouds experiencing the greatest acceleration are also matter-bounded clouds, which don't contribute to the reverberation response, but still contribute to the average line profile.  Additionally, allowing the cloud mass to vary, such as varying cloud mass, could produce less boxy profiles. Finally, a BLR model with combined velocity fields, with non-radial components contributing to the LOS the velocity (such as turbulence and/or Keplerian motion) will tend to produce more Gaussian or Lorentz-like line profiles that would be more representative of observed BELs.            

\subsection{Representation of Real Broad Emission Line Responses}
\label{sec:dis:real}

Utilizing simple recombination theory (Equation~\ref{eq:lum}) for the cloud luminosity is a reasonably good approximation for H$\alpha$ emission.  Roughly speaking, the H$\alpha$ flux emitted per unit area by a given cloud is proportional to the column density of ionized gas, $N_s(r)$.  This in turn depends on $U(r,t^\prime)$ and therefore responds to a changing ionizing radiation field. We can also make general statements about the emissivity distribution for other broad lines.  
For instance, the high-ionization emission lines, such as  CIV$\lambda\lambda\,1551,\,1548\,$\AA, are mainly emitted near $R_{in}$ where the ionizing radiation field is strongest \citep{Osterbrock1986Emission-LineQSOs}.  The response of CIV would peak at a short time delay, then rapidly drop off with time delay.  Therefore, we'd expect the response behavior of high-ionization lines like CIV to be somewhat similar to models with small values of $s$ and $p$.  Conversely, low-ionization lines, such as MgII$\lambda\lambda\,2795,\,2803\,$\AA, have an emissivity distribution that varies more slowly with radius and so can be approximated by specifying larger $s$ and $p$ values.  The power-law approximation of $\varepsilon(r)$ includes the $s$ and $p$ parameters and should provide a general guide to the dependence of a given line's reverberation response to $U(r,t^\prime)$, $n(r)$, and $N_{col}(r)$.  However, the emissivity distributions for true emission lines in BLRs are certainly more complex functions than implied by Equation~\ref{eq:emiss}.  For a more sophisticated calculation of the line emission fluxes, the next version of BELMAC will include photoionization model grids for selected BELs.       

When the $elf$ is used, Equation~\ref{eq:elf}, it provides a reasonable approximation for anisotropic line emission from a distribution of clouds and the ACE models should be generally representative of emission lines that are optically thick.  For example, ACE models can be utilized to approximate the response of Ly$\alpha$ that is emitted in the ionized part of the cloud and escapes almost entirely through the illuminated face, but has a very low probability of escaping through the non-illuminated face.  ICE models represent lines that are optically thin, such as CIV, which can escape isotropically if the cloud contains no dust.  Furthermore, we assume that a cloud which becomes matter-bounded also becomes effectively optically thin, allowing the line emission to escape isotropically.  We could expect dust to survive in the partially ionized and neutral interiors of BLR clouds \citep{Baskin2018DustNuclei}.  Dust is not explicitly included in the current version of BELMAC, but ACE, modeled with the $elf$ parameter, could also be used to approximate the effects of dust extinction in BLR clouds.    

\subsection{Inferring Black Hole Mass from the Reverberation Lag}
\label{sec:dis:MBH}

As discussed in Section~\ref{sec:intro} an important motivation for BLR reverberation mapping is to determine $R_{BLR}$ and hence $M_\bullet$ via the virial theorem (Equation~\ref{eq:BHmass}).  The BLR models presented Section~\ref{sec:results} show that the RWD is not always equivalent to the LWR. Matter-bounded and anisotropically emitting clouds have a significant affect on the response-weighted lag, $t^\prime_{RW}$, which could lead to overestimates of the SMBH mass.  If we assume the velocity dispersion is determined exactly and assume $ct^\prime_{RW}=R_{LW}=R_{BLR}$ ($2R_d$LWR $=R_{LW}$; see Section~\ref{sec:intro:RM}) then, 
\begin{equation} 
    \frac{M_\bullet}{fM_{\bullet,\,vir}} = \frac{\mathrm{LWR}}{\mathrm{RWD}}
\end{equation}
where $M_\bullet$ is the true black hole mass and $M_{\bullet,\,vir}$ is the mass inferred from reverberation mapping.  Since we are assuming $2R_d$LWR $=R_{BLR}$, if LWR/RWD $<1$ then RWD is overestimating LWR and therefore also $M_\bullet$. This is the case for all of the models with matter-bounded clouds and/or ACE presented in Figures~\ref{fig:sphere:LWR} and \ref{fig:disk:LWR}.                 

\begin{figure}
\centering
\includegraphics[trim=0 110 0 110, clip, width=\textwidth]{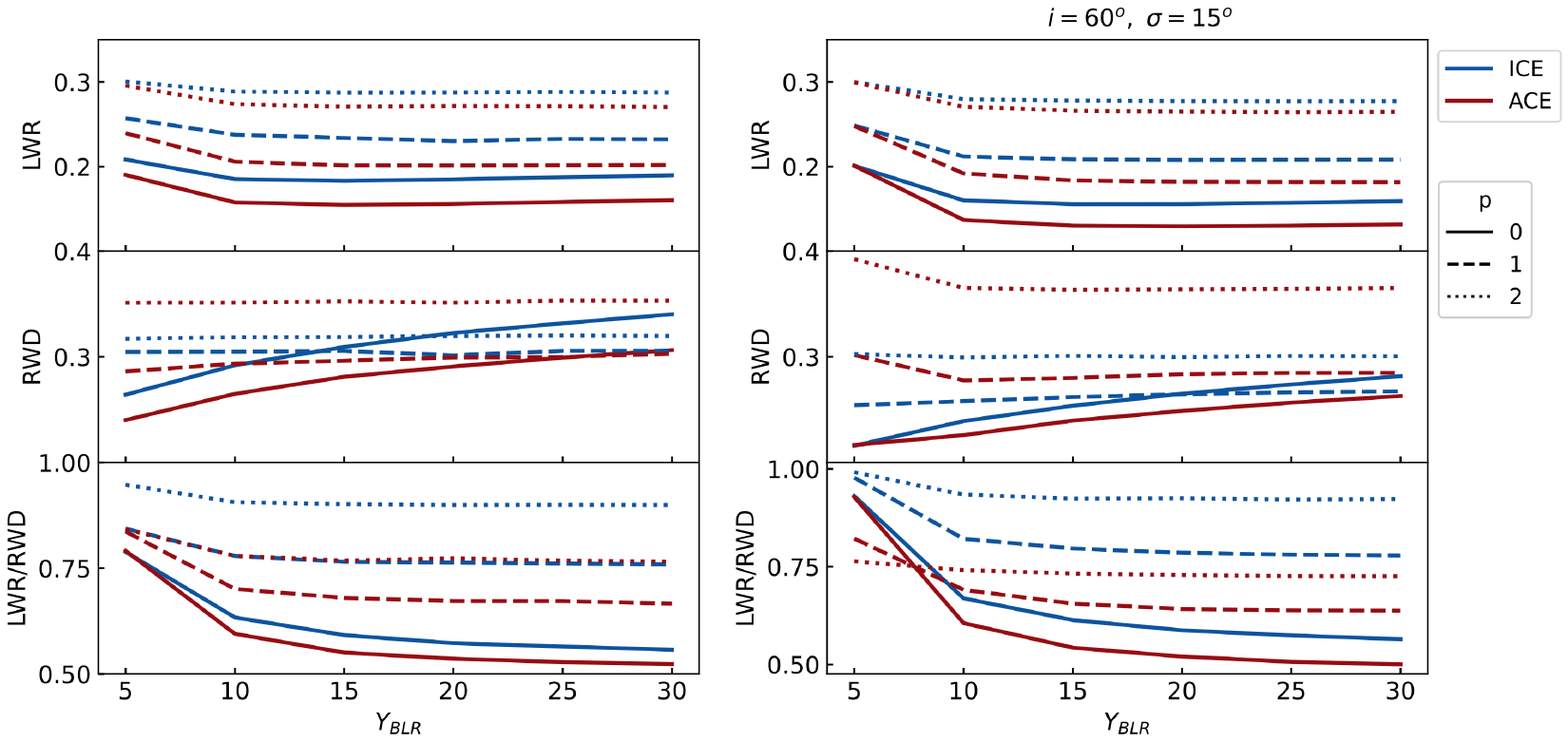}
\caption{The LWR, RWD, and LWR/RWD for (left) spherical and (right) disk-like ($\sigma=15^o,~i=60^o$) BLR models, but for only models containing matter-bounded clouds ($s=0$).  The blue and red lines represent ICE and ACE BLR models, respectively.}
\label{fig:LWR:s0}
\end{figure}

Figure~\ref{fig:LWR:s0} shows the LWRs and RWDs for the spherical and disk BLR models that include matter-bounded clouds from Figures~\ref{fig:sphere:LWR} and \ref{fig:disk:LWR}.  The bottom panels show the RWD can be $\sim1.1\times$ LWR, for the smallest BLRs, to nearly $2\times$ LWR for $Y_{BLR} \gtrsim 20$. ACE also results in LWR/RWD $<1$, even when 100$\%$ of the clouds are radiation-bounded. Figures~\ref{fig:sphere:LWR} and \ref{fig:disk:LWR} show that RWD $\approx 1.2 - 1.5\times$LWR for ACE models with no matter-bounded clouds ($s=-1, -2$).  As discussed above, ACE is a fair representation of optically thick emission lines and the possible extinction by dust embedded in the cooler regions of the clouds.  Therefore, in the sample of BLR models explored here, $R_{BLR}$ is correctly determined from $t_{RW}$ only when all the clouds in the BLR ensemble are radiation-bounded \textit{and} emitting isotropically.  Given the conditions in real BLR, this scenario is unlikely to occur for Balmer lines, which tend to be optically thick \citep{Osterbrock1986Emission-LineQSOs}. This suggests the lag measured from cross-correlation does not relate to $R_{BLR}$ as straightforwardly as previously assumed.  However, we will be revisiting these models with a version of BELMAC that implements photoionization models in forthcoming work.  

% LWR, RWD for s=0

\subsection{Comparison to TORMAC}
\label{sec:dis:vs tormac}

Although TORMAC was designed to simulate reverberation of the torus dust emission in AGN, it is worth discussing similarities and differences between TORMAC and BELMAC.  

TORMAC includes physics potentially important for both the torus and the BLR; cloud occultation, cloud shadowing, and the effect of cloud orientation.  A cloud is occulted and its emission is attenuated when another cloud is positioned along the LOS between the former cloud and the observer.  The strength of the attenuation depends on the number of occulting clouds and their optical depths.  Cloud shadowing occurs when a cloud is positioned within the shadow of another cloud intervening between it and the illuminating source.  A cloud without direct exposure to the source is indirectly heated by the diffuse emission from neighboring clouds and emits isotropically \citep{Almeyda2017ModelingIllumination,Almeyda2020ModelingFunctions}.  As a general point, we expect cloud occultation and shadowing to be less important for the BLR's gas clouds, even if the clouds themselves are optically thick for a given line, because the BLR as a whole is presumed to be optically thin (i.e., $C_f<1$).  Therefore, these two effects are not included in the current version of BELMAC, but could be incorporated in later versions.  Cloud emission anisotropy, on the other hand, is expected to be important in some situations and is incorporated in BELMAC via the $elf$ parameter (Equation~\ref{eq:elf}).  This is based on the orientation-dependent fraction of the cloud's illuminated surface that is viewed by the observer.  TORMAC includes a more sophisticated implementation of cloud orientation effects than BELMAC's $elf$ parameter.  The anisotropic emission of a cloud in TORMAC, adopted from the clumpy torus model by \citet{Nenkova2008AGNMEDIA}, is wavelength dependent.  In this model, the spectrum of a cloud is approximated by averaging over the emission from many slabs of gas at varying orientation angles relative to the central illuminating source  \citep{Almeyda2017ModelingIllumination,Almeyda2020ModelingFunctions}. 

TORMAC also includes anisotropic illumination of the torus.  Edge darkening of the AGN's accretion disk causes the illuminating radiation field to be anisotropic.  As a result, the dust sublimation radius is not constant, but a function of polar angle and the dust can exist closer to the source near the midplane than towards the pole.  For dust clouds arranged in a sphere and anisotropically illuminated, the cross-section of the inner dust free cavity would have a figure-8 shape \citep{Almeyda2017ModelingIllumination,Baskin2018DustNuclei,Kudoh2023MultiphaseOutflow}.  Although we do not explore anisotropic illumination here, it is included in BELMAC.  For the BLR, this effect will result in both the ionization parameter (Equation~\ref{eq:U}) and the outer boundary, the dust sublimation radius, becoming functions of polar angle, $\theta$.              

\subsection{Other Reverberation Mapping Modeling Codes}

\citet{Mangham2019DoRegion} categorizes 2 methods for deciphering the structure and kinematics of the BLR from velocity-resolved reverberation mapping data.  The `inverse' method constructs the 2DRF from data. For example, the Maximum Entropy Method (MEM) is a deconvolution technique utilized by the echo mapping code, \textsc{Memecho}, to recover the velocity-resolved transfer function from observations and fit reverberation mapping data \citep{Horne1994EchoSolutions}.  However, it is still necessary to interpret the results to determine the geometry and dynamics. 

The other method is forward modeling, which is the approach used by BELMAC, \textsc{Caramel} (Code for AGN Reverberation and Modeling of Emission Lines) by \citet{Pancoast2011GEOMETRICDATA}, and \textsc{BRAINS} by \citet{Li2013ADATA}.  BELMAC and \textsc{Caramel} both create a 3D ensemble of clouds, however, \textsc{Caramel} utilizes a more prescriptive parameterization of BLR properties.  For example, \textsc{Caramel} treats the clouds as particles that are representative of a gas density field and approximates the responding emission by setting the emissivity as a radial power-law function \citep{Williams2022CARAMEL-gas:Gas}.  In contrast, BELMAC models the BLR as an ensemble of discrete clouds, whose line emission is calculated using nebular physics.  We used hydrogen recombination theory for the models presented here, but a large grid of photoionization models will be used in the forthcoming version.  \textsc{Caramel} includes a parameter estimation procedure to recover the best matching model for a time series dataset.  This capability is currently not implemented in BELMAC, but in the future we will include a parameter estimation procedure. This will make BELMAC a valuable tool for BLR reverberation mapping modeling that will compliment existing modeling codes such as {\sc Caramel}, while offering different capabilities.
  
The BLR Reverberation-mapping Analysis In AGNs with Nested Sampling (\textsc{BRAINS}) code, developed by \citet{Li2013ADATA}, is based on \textsc{Caramel} and employs similar methodology. However, \textsc{BRAINS} allows the BLR emission to respond non-linearly and additionally, \citet{Li2018Supermassive142} added a 2-zone BLR geometry, where each zone may have a different cloud distribution, which could be used to model a disk-like BLR with a wind.  BELMAC may also be used to model a 2-zone BLR, with different gas density and cloud distributions, velocity fields, and radial sizes, but we leave this discussion for a future paper.  Lastly, \textsc{BRAINS} also models the velocity-resolved spectroastrometric response \citep[][and references therein]{Li2023SpectroastrometricRegions}.  This is not currently a feature in BELMAC, but could be implemented with straightforward modifications.

\section{Conclusion} \label{sec:concl}

We have presented a parameter exploration of velocity-resolved, reverberation response functions (numerical approximations of the transfer function) of the BLR using BELMAC.  The main purpose of BELMAC is to serve as a versatile forward modeling tool for analyzing BLR reverberation mapping data.  BELMAC is designed for flexibility, allowing various different combinations of geometry and dynamical models to be incorporated.  Here, we focused on 2 distinct BLR models; a spherical shell with a radiation-pressure driven outflow and a thin disk with Keplerian motion.  In the models presented in this paper, the broad H$\alpha$ line emission is calculated using an analytical prescription based on recombination theory.  The main results of this paper a summarized below.           

\begin{enumerate} 

\item The $\tau-v^{||}$ relationship is linear for spherical radial outflows, and elliptical for rotating disks.  This results in wedge-shaped and bell-shaped velocity-delay maps (2DRFs), respectively, as already described in earlier work \citep[e.g.,][]{Perez1992TheFunctions,Horne1994EchoSolutions,Pancoast2014ModellingResults}.  

\item For models with isotropic cloud emission (ICE) and no matter-bounded clouds, the 1DRFs are flat-topped for spherical shell BLRs and double-peaked for disk-like BLRs, for delays less than the light-crossing time of the inner cavity ($\tau\leq1/Y_{BLR}$).  For $\tau>1/Y_{BLR}$, the 1DRFs for both shells and disks decay as $\tau$ increases, but the shape of the decay depends on the gas density profile and how the clouds are distributed.  Whether for a sphere or a disk, the 1DRFs become more extended as $s$ decreases and $p$ increases ($\varepsilon_V(r)$ decreases more slowly with $r$).  Additionally for the disk models, the time delay difference between the double-peaks will reduce with decreasing inclination.      

\item  For models with ICE and no matter-bounded clouds, the time-averaged line profiles are flat-topped for spherical BLR models with radial outflows and double-peaked for rotating disk models.  In spherical, radial outflow models, the response begins in the blue-wing of the line profile then propagates to the red-wing.  Therefore, the RMS profiles tend to be skewed blueward, but the time-averaged line profiles are symmetric. In the rotating disk models, the response begins in the profile core, then propagates to the maximum velocity, $\pm v^{||}_{Kep}$, in the wings, and finally moves symmetrically back towards the core.  The averaged line profiles have symmetrical double-peaks, which have separations dependent on the disk's inclination.  The RMS profiles also have symmetrical double-peaks, however, at large inclinations, the response amplitude at $v^{||}_{Kep}\approx 0$\,km\,s$^{-1}$ and $\tau\approx0$ is strong enough that the RMS spectra are triple-peaked.   

\item We assume that recombination theory provides a reasonable estimation of the broad H$\alpha$ response.  Considering lines other than the hydrogen Balmer Series, the models with ICE can be considered to represent emission lines that are likely to be optically thin, such as CIV, and likewise, the ACE models represent lines, such as Ly$\alpha$, that are likely to be optically thick.    

\item  Both matter-bounded clouds and ACE can affect the response quite dramatically because these effects suppress the response of the inner clouds, or the near-side clouds, respectively.  Both effects cause the 1DRF to become more extended in time delay.  The models with matter-bounded clouds respond as if the BLR has an effective inner radius that is greater than the true value of $R_{in}$.  In ACE models, the 1DRFs tend to peak at delays longer than $1/Y_{BLR}$.  In radial outflow models, ACE causes redward asymmetric line profiles.  For rotating disks, the line and RMS profiles decrease in amplitude by half, but ACE doesn't affect the shapes.  However, matter-bounded clouds emit isotropically and therefore partly mitigate the profile asymmetry and decreased amplitude.     

\item The luminosity-weighted radius is equivalent to the response-weighted delay when the BLR is comprised entirely of radiation-bounded, isotropically emitting clouds.  However, the RWD can significantly exceed the LWR when matter-bounded clouds are present or when clouds emit anisotropically.  These effects in combination lead to over-estimations of $R_{BLR}$ by $\sim10-100\%$ implying that measured cross-correlation lags may also over-estimate $R_{BLR}$ in such circumstances.   

\end{enumerate}

We will use photoionization models grids to explore multi-emission line responses in a forthcoming paper.  More complex combinations of geometries and velocity fields, such as a rotating disk with a biconical wind, will also be presented in a later paper. In the future, BELMAC will include a parameter estimation procedure for matching response models to BLR reverberation mapping data.  BELMAC will compliment existing BLR reverberation modeling codes such as {\sc Caramel, Memecho, and BRAINS}.

\begin{acknowledgments}

This paper is based upon work supported by the National Science Foundation under AARG/Grant Number 2009508.  

\end{acknowledgments}


\vspace{5mm}

\section{Author publication charges} \label{sec:pubcharge}

%Finally some information about the AAS Journal's publication charges. In April 2011 the traditional way of calculating author charges based on the number of printed pages was changed.  The reason for the change was due to a recognition of the growing number of article items that could not be represented in print. Now author charges are determined by a number of digital ``quanta''.  A single quantum is 350 words, one figure, one table, and one enhanced digital item.  For the latter this includes machine readable tables, figure sets, animations, and interactive figures.  The current cost for the different quanta types is available at \url{https://journals.aas.org/article-charges-and-copyright/#author_publication_charges}. Authors may use the ApJL length calculator to get a {\tt rough} estimate of the number of word and float quanta in their manuscript. The calculator is located at \url{https://authortools.aas.org/ApJL/betacountwords.html}.


\bibliography{references}
\bibliographystyle{aasjournal}



\end{document}